\documentclass[12pt]{article}
\usepackage{graphics,epsfig}

\def\dalemb#1#2{{\vbox{\hrule height .#2pt
        \hbox{\vrule width.#2pt height#1pt \kern#1pt
                \vrule width.#2pt}
        \hrule height.#2pt}}}

\let\a=\alpha  \let\g=\gamma \let\d=\delta \let\e=\epsilon
  \let\th=\theta  \let\k=\kappa
\let\l=\lambda \let\m=\mu \let\n=\nu \let\x=\xi  
\let\s=\sigma \let\t=\tau    
 
\let\w=\omega      \let\G=\Gamma \let\D=\Delta \let\Th=\Theta 
\let\X=\Xi  \let\S=\Sigma  \let\Y=\Psi
 
\let\la=\label \let\ci=\cite 
  
\def\nn{\nonumber} \def\bd{\begin{document}} \def\ed{\end{document}}
\def\ds{\documentstyle} \let\fr=\frac \let\bl=\bigl \let\br=\bigr
\let\Br=\Bigr \let\Bl=\Bigl
\let\bm=\bibitem
\let\na=\nabla
\def\tU{{\widetilde U}}
\let\pa=\partial \let\ov=\overline
\def\ie{{\it i.e.\ }}
\newcommand{\be}{\begin{equation}}
\newcommand{\ee}{\end{equation}}
\def\ba{\begin{array}}
\def\ea{\end{array}}
\def\ft#1#2{{\textstyle{{\scriptstyle #1}\over {\scriptstyle #2}}}}
\def\fft#1#2{{#1 \over #2}}
\def\F#1#2{{ F_{#1}^{(#2)} }}
\def\cF#1#2{{ {\cal F}_{#1}^{(#2)} }}

\def\R{{\bf R}}
\def\sst#1{{\scriptscriptstyle #1}}
\def\oneone{\rlap 1\mkern4mu{\rm l}}
\def\e7{E_{7(+7)}}
\def\td{\tilde}
\def\wtd{\widetilde}
\def\im{{\rm i}}
\def\bog{Bogomol'nyi\ }
\newcommand{\ho}[1]{$\, ^{#1}$}
\newcommand{\hoch}[1]{$\, ^{#1}$}
\newcommand{\bea}{\begin{eqnarray}}
\newcommand{\eea}{\end{eqnarray}}
\newcommand{\ra}{\rightarrow}
\newcommand{\lra}{\longrightarrow}
\newcommand{\Lra}{\Leftrightarrow}
\newcommand{\ap}{\alpha^\prime}
\newcommand{\bp}{\tilde \beta^\prime}
\newcommand{\cB}{{\cal B}}
\newcommand{\cO}{{\cal O}}
\newcommand{\vecx}{\vec{x}}
\newcommand{\vecy}{\vec{y}}
\newcommand{\vecp}{\vec{p}}
\newcommand{\vecq}{\vec{q}}
\newcommand{\tr}{{\rm tr} }
\newcommand{\Tr}{{\rm Tr} }
\newcommand{\NP}{Nucl. Phys. }

\newcommand{\cL}{{\cal L}}
\newcommand{\cA}{{\cal A}}
\newcommand{\cD}{{\cal D}}

\def\sst#1{{\scriptscriptstyle #1}}
\def\0{{\sst{(0)}}}
\def\1{{\sst{(1)}}}
\def\2{{\sst{(2)}}}
\def\3{{\sst{(3)}}}
\def\4{{\sst{(4)}}}
\def\5{{\sst{(5)}}}
\def\6{{\sst{(6)}}}
\def\7{{\sst{(7)}}}
\def\8{{\sst{(8)}}}
\def\ve{\varepsilon}
\def\vf{\varphi}
\def\F{\Phi}
\def\wg{\wedge}

\newcommand{\tamphys}{\it 
}

\newcommand{\auth}{AUTHORS}

\def\thb{\bar{\theta}}
\def\Thb{\bar{\Theta}}
\def\barp{\bar{p}}
\def\barq{\bar{q}}
\def\barc{\bar{c}}
\def\bard{\bar{d}}
\def\e{\epsilon}

\def \bi{\bibitem}
\def \la {\label}

\def \l {\lambda}
\def\foot{\footnote}
\def \tl  {{\tilde \l}}
\def \sql {{\sqrt \l}}
\def \adss {$AdS_5 \times S^5$\ }
\newcommand{\rf}[1]{(\ref{#1})}
\def \ov {\over}

\def\th{\theta}
\def\Th{\Theta}
\def\vth{\vartheta}
\def\btheta{{\bar\theta}}
\def\ttheta{{{\tilde\theta}}}
\def\bttheta{{{\bar\ttheta}}}
\def\vth{\vartheta}

\def\ra{\rightarrow}
\def\N{{\cal N}}
\def\F{{\cal F}}
\def\uM{\underline{M}}
\def\uN{\underline{N}}
\def\uP{\underline{P}}
\def\cc{\circ}
\def\eqv{\equiv}

\def\ni{\noindent}

\def\Ep{E^{{}^{(+)}}}
\def\Em{E^{{}^{(-)}}}

\def\Mp{M^{{}^{(+)}}}
\def\Mm{M^{{}^{(-)}}}

\def \ha{{1\ov 2}}

\def\r{{\rm r}}

\def\Y{{\rm Y}}
\def\X{{\rm X}}
\def\tY{\tilde{\rm Y}}
\def\tX{\tilde{\rm X}}
\def\dY{\dot{\rm Y}}
\def\dX{\dot{\rm X}}

\def \J {\mathcal{J}}
\def \del {\partial}

\def\dF{\dot{F}}
\def\dG{\dot{G}}
\def\df{\dot{f}}
\def \E {{\cal E}}
\def \S {{\cal S}}
\def \J {{\cal J}}

\def\ms{\mathcal{S}}
\def\mj{\mathcal{J}}
\def\soj{\fr{\ms}{\mj}}
\def \R {{\bf R}}
\def \om {\omega}
\def \bE {\bar E}
\def \x {{\cal X}}

\begin{document}
\overfullrule=0pt
\parskip=2pt
\parindent=12pt
\headheight=0in \headsep=0in \topmargin=0in
\oddsidemargin=0in

\vspace{ -3cm}
\thispagestyle{empty}

\begin{center}

{\Large\bf
Spinning strings in $AdS_5 \times S^5$:
 \\
 \vspace{0.1cm}
 one-loop  correction  to energy in $SL(2)$ sector}

 \vspace{.5cm} { I.Y. Park\footnote{ipark@mps.ohio-state.edu },
 A. Tirziu\footnote{tirziu@mps.ohio-state.edu }  and A.A.
 Tseytlin\footnote{Also at Imperial College London
 and  Lebedev  Institute, Moscow.
 }}\\
 \vskip 0.3cm

{Department of Physics, The Ohio State University,\\
Columbus, OH 43210, USA   }

\end{center}

\def \bi{\bibitem}
\def \la {\label}

\def \l {\lambda}
\def\foot{\footnote}
\def \tl  {{\tilde \l}}
\def \sql {{\sqrt \l}}
\def \adss {$AdS_5 \times S^5$\ }
\def \ov {\over}

\def \varpi {{\rm w}}

 \vspace{0.1cm}

 \begin{abstract}
We consider  a circular string with
spin $S$  in $AdS_5$
 wrapped around  big circle of $S^5$ and carrying also
 momentum $J$. The corresponding $N=4$  SYM operator
  belongs to the $SL(2)$ sector, i.e. has  tr$(D^S Z^J)+...$
  structure. The leading  large $J$ term in its 1-loop
anomalous dimension can be computed  using Bethe ansatz for
 the $SL(2)$ spin chain and was  previously  found to match
 the leading   term in the classical string energy.
The string  solution is stable at large $J$,  and the
Lagrangian for string fluctuations
has constant coefficients, so that the
1-loop string correction  to the   energy $E_1$
is given simply by the sum of characteristic frequencies.
Curiously, we find that  the leading term in the
zero-mode part of $E_1$
is  the same   as a  $1/J$ correction to the one-loop
anomalous dimension
 on the gauge theory (spin chain)
  side that was  found  in hep-th/0410105. However,
the  contribution of non-zero string  modes does not
vanish.
We also discuss the ``fast string''  expansion of the
classical string action which coincides  with the coherent
state action of the $SL(2)$ spin chain at the first
order in $\l$, and  extend this expansion to higher orders
 clarifying   the role of the $S^5$ winding number.

\end{abstract}
\newpage

\setcounter{equation}{0}
\setcounter{footnote}{0}
\setcounter{section}{0}

\renewcommand{\theequation}{1.\arabic{equation}}
 \setcounter{equation}{0}

\section{Introduction}

Study  of semiclassical  rotating strings in \adss
which extended earlier work \ci{bmn,gkp} to cases
with several large  angular momenta
have  led to   interesting developments
in  understanding and checking the AdS/CFT  duality.
The classical energy of such strings    has
 ``regular'' expansion \ci{ft1,ft2}  in
  the effective coupling $\tl = {\l \ov J^2}$, i.e.
$E= J ( 1 +  c_1 { \lambda \ov J^2} + c_2 { \lambda^2 \ov J^4}+
 ...) $, where
$J$ is total angular momentum and $\l$ is the
 square of string tension or the `t Hooft coupling on the
 SYM side. This
 prompted  a
possibility \ci{ft2} of direct comparison with perturbative
anomalous dimensions of the corresponding single-trace operators
in  $\N=4$ SYM theory.

Indeed, the precise agreement was
established at the first two leading orders in $\l$
\ci{bmsz,bfst,mez,ss,minah}  (for reviews and further references see
\ci{ts1,ts2,beis,kmmz,zare}). This  agreement was found to
break down (as in the near-BMN \ci{ft1,par}
 case \ci{callan}) at
the next $\l^3$ order \ci{ss,minah}.
 A natural explanation   is
an  order-of-limits problem \ci{ss,bds,afs,s}:
on the string side one first takes
$J$ to be large to suppress quantum ($\a'$) string corrections
and then expands $E/J$ in $\tl = {\l \ov J^2}$ (so that
$\l$ is effectively  large), while on the SYM side one uses
perturbation theory  in $\l$  and then
expands each loop correction to anomalous dimensions in  $1/J$.
The exact  agreement at the first two leading orders appears to be due
 to a
special  structure of the one- and two-loop dilatation  operator
in $\N=4$ SYM theory \ci{mz,bks,b2} and
should be essentially a consequence of
 large supersymmetry of the theory.

To understand how one may  still preserve  the
 equality between the exact string energies $E(J, \l)$
and the exact SYM anomalous dimensions $\Delta(\l,J)$,
i.e. to explain the  interpolation between the perturbative string and the perturbative SYM
expressions
 it is important to go beyond the classical string theory and
 compare the 1-loop corrections to string energies
 to subleading in  $1/J$ terms in the SYM anomalous dimensions.
This is also important in order  try to provide more
data  for checking
a non-perturbative Bethe ansatz proposal of  \ci{afs}.

 On the string side, the computation of 1-loop corrections
 to classical string energies
 is relatively  straightforward for a special class of
 ``homogeneous'' circular strings \ci{ft2,art}
 for which the fluctuation Lagrangian  has constant
 coefficients \ci{ft3,art}.
On the SYM side, finding  the corresponding  subleading $1/ J^2$
terms in the  one-loop anomalous dimensions amounts to  computing
corrections to the thermodynamic limit of the Bethe ansatz equations
and was previously done in the $SU(2)$ sector \ci{lz}   only  for a
special state corresponding to a circular string rotating  in $S^5$
with two equal angular momenta \ci{ft2}.
The latter  solution is, however, unstable \ci{ft2,ft3},   so that
the 1-loop correction to its energy    contains formally an
imaginary part. Then  a mismatch between  its real part
 \ci{fpt} and
a  real SYM expression of \ci{lz}  may   be viewed
as an inconclusive evidence of a disagreement.

Here we shall  perform   a 1-loop string
computation very similar to the one carried out in
\ci{ft3,fpt} but for a {\it stable}
  circular 2-spin  string solution found in  \ci{art}.
  In conformal gauge, it is represented
by a  string of fixed radius (wound $k$ times)
lying on a plane
in $AdS_5$  and
 rotating along itself so that it carries one component $S$
 of spin in $AdS_5$.  The string is also wound ($m$ times) around a big circle
 of $S^5$  and  is rotating  along itself with  momentum $J$
 (the ``level matching'' constraint implies that $k S + m J=0$).
  The fast enough rotation  stabilizes this solution.
 This configuration
 may be visualized as a circular spiral.\foot{This is
 a  fixed time
  profile of the string, so
  a more appropriate name would be a ``spiral string''.}
  It   is an \adss analog of
 a   closed  string  in flat  $\R_t \times \R^{2} \times S^1$
 space-time
 which is wrapped $k$ times on a constant-radius
 circle in $\R^2$  and also
 $m$ times on $S^1$ and rotating along itself in each circle
 (alternatively, the flat-space solution
   may be viewed as a left-moving wave along a
 circular string  wrapped on $S^1$  and it thus represents
  a BPS
 state; the  \adss solution is no longer BPS).

The classical energy of this $(S,J)$   string
solution  has the following expansion at
 large $J$ with fixed $S\ov J$  \ci{art}:
\be \la{cla}
E_0 =  J + S + { \l k^2 \ov 2 J} { S\ov J} ( 1 + { S\ov J})
+ O( { \l^2  \ov  J^3}) \ .
\ee
The corresponding SYM operator belongs
 to the closed  $SL(2,R)$
 sector in which the 1-loop  SYM
 dilatation  operator is the same
 as the  Hamiltonian
 of the $SL(2)$ Heisenberg spin chain with
  length $L=J$ \ci{bs}, i.e.
it  has the structure
 tr$ (D^S Z^J)+ ...$,  where
 one is to sum over orderings of $D$'s and $Z$'s
 to get an eigen-state of the dilatation operator
 ($D= D_1 + i D_2$ is a covariant derivative in $\R^4$
 and  $Z= \phi_1  + i \phi_2$ is a  chiral combination of
 SYM scalars).  The  analysis  of the thermodynamic limit
 of the 1-loop $SL(2)$ spin chain Bethe ansatz equations
 showed \ci{kz} that there is indeed  a SYM state  whose
 1-loop anomalous dimension $\Delta (S,J; \l)$
    matches precisely the order $\l$ term in \rf{cla}.
 A  subleading  $1/ J^2$ term in the  1-loop part of $\D$
  can be found \ci{kz} as in  the $SU(2)$ sector \ci{lz} from
  the  correction to the thermodinamic limit of the Bethe ansatz
  equations. It  turned out  to be given simply by the leading-order
   term
  multiplied by $-{1 / J}$, i.e. \ci{kz}
  \be \la{sym}
 \Delta =
 \l  {  k^2 \ov 2 } { S\ov J} ( 1 + { S\ov J})
 \bigg[{ 1 \ov J}  -  { 1 \ov J^2 } + O( { 1  \ov  J^3}) \bigg]
 + O(\l^2) \ . \ee
Our aim below will be to compute the 1-loop
string correction $E_1$
to the  energy \rf{cla} of the
circular $(S,J)$ string solution.
We shall find that, as expected \ci{ft3}, the 1-loop correction is
suppressed compared to the classical expression \rf{cla} by an extra
power of $J$,
  and will compare
the coefficient of the
order ${\l \ov J^2}$ term in $E_1$  to the
 coefficient of the ${ \l \ov J^2 }$ term on the SYM
 side  \rf{sym}.

 As in the $SU(2)$  case considered in \ci{fpt}, we shall
conclude
 (in sect.4 below)  that there is an apparent
 disagreement between the subleading in $1/ J$
 string result of this paper  and
the spin chain  result of \ci{lz,kz}.
A
 curious observation is that
  the agreement would hold if we  were to keep only the
 contributions of the
 zero modes of the  string fluctuations, i.e. to
 consider a  ``homogeneous''    approximation
 in which the  string fluctuations
 depend only on $\tau$ but not on  $\sigma$.

 One may think, of course, that there is
 no a priori reason to expect the exact
 agreement, and, as  in \ci{ss,bds,fpt},  any
  disagreement should be attributed to
  the different order of limits
 taken on the gauge theory and the string theory sides of the
 AdS/CFT duality.\foot{There may be  potential
  subtleties related to how one takes limits in the
  $(\l, J)$ parameter space \ci{bds}. In the cases of  subleading $1/J^2$  corrections to the BMN
  point-like state
  and the $1/J $ corrections to the spinning strings the
  disagreements could, in principle, start already at
  order $\l$ \ci{fpt,s}.
Still, one could  hope,
  by analogy with what was found for
  the large $J$ correspondence,
  that disagreements should
 start only at order $\l^3$
 (e.g., due to a specific  structure of the 3-loop SYM dilatation
  operator).}
 In particular, the  spin chain Bethe ansatz and the string theory
  computations of
 the subleading $1/J$ corrections are not directly
 related:
 while the  Bethe ansatz approach
 does reproduce part of the spectrum  of the bosonic
 string fluctuations near a given semiclassical
 solution \ci{bmsz,kz},\foot{There are two type of
 string fluctuations near circular spinning strings
  (see \ci{ft2,ft3} and below):
 whose frequencies
 scale at large $\J = { J \ov \sqrt \l}$
 as  (i)  $ \omega^{(1)}
 \sim \J + { a\ov  \J }$  and as  (ii) $\omega^{(2)}\sim
 { b\ov  \J }$
 (the corresponding energies are $ \delta E
  \sim { \om \ov \J}$). Fermionic frequencies
  belong to the first type.
  The first (``BMN'') type of   fluctuations  correspond
  to deformations away from a particular ($SU(2)$ or $SL(2)$)
  sector, and are not ``seen'' directly in the Bethe ansatz
  equations for the corresponding spin chain.
   The second type of fluctuations correspond
  to deformations  within a given sector  and are captured by
  the spin chain \ci{bmsz,frey,minah} and the
  corresponding coherent state (``Landau-Lifshitz'')
   action \ci{kru,krt,hl1}.}
  the string 1-loop computation involves
  summing over {\it all}  bosonic {\it and}
 fermionic  modes.\foot{In the Bethe ansatz
 approach fluctuations near a given
 semiclassical state described by a
 smooth Bethe root distribution are obtained
 by pulling one Bethe
 root out of a continuous distribution \ci{bmsz}. Summing over fluctuations would
 look like averaging over a set of Bethe states, which is not what was
 done in \ci{lz}  to compute the $1/J$ correction.
 We are grateful to K. Zarembo  for this remark.}

Before turning to the computation of the one-loop
correction to the energy
of the circular   solution we shall start  (in sect. 2) with
a discussion  of   the ``fast string'' limit of the classical
string action
in the   $SL(2)$ or $(S,J)$ sector.
The  comparison of the reduced sigma model
appearing in the  ``fast string'' limit of the
 string action to the coherent-state
(``Landau-Lifshitz'')
action on the ferromagnetic spin chain side
provided a simple and    universal way
of   matching the   string and gauge theory
semiclassical states as well as  near-by fluctuations
\ci{kru,krt,hl1,st,mik12,kt,ital,mik3,hl2}.\foot{A
 complementary
 approach  demonstrating  general agreement
 between the string and gauge theory semiclassical states
  at the two leading orders in $\l$
 is based  on directly relating the thermodynamic
 limit of the Bethe ansatz
equations to the integral equation
representing the  general integrability-based
solution of the corresponding sector of
the classical \adss sigma model
\ci{kmmz,kz,as,bksa,af,sak}.}
In the $SL(2)$  sector that was done in \ci{st} and also
in \ci{ital}.
  Below we will  extend
 the  derivation  of the string action limit
 in \ci{st} to the next  order of expansion in
  $\l \ov J^2 $  and will
  also include  the general case of string
  configurations  with
non-zero winding number $m$ in the $S^1 \subset S^5$ direction.
The spin-chain coherent state action cannot depend on  $m$,
and, indeed,  $m$ does  not  appear also
in the relevant limit of the string sigma model action; instead,
it  classifies  particular solutions of the
resulting  ``Landau-Lifshitz'' equations.
In particular,
we shall explain (in sect. 3)  how
 the circular $(S,J)$ solution of \ci{art} which has
  $m\not=0$
can be obtained
 from the $SL(2)$
``Landau-Lifshitz'' equations (order
by order in $\sql\ov J $).
Closely related aspects of the general $m\not=0$
sigma model solutions in the
 $SL(2)$ sector   and their relation  to
  the Bethe root distributions
 on the spin chain side were discussed in \ci{kz}.

The circular string  state of $SL(2)$ sector   discussed
in this paper may  be of interest  also in a
more general context of
gauge--string duality.
Similar solutions exist in
any space  with $AdS_3 \times S^1$ part, e.g.,
in $AdS_5 \times S^1$ which is  relevant for a
description of less supersymmetric    4-d conformal theories
\ci{poly}. Also, the corresponding operators in the  $SL(2)$ sector
of $\N=4$ SYM have  direct  connection
to operators  appearing in pure YM
  or $\N=1$ SYM  theory \ci{bgk}.

\bigskip

We shall start in section 2 with a discussion of large
$J$ expansion of the
classical string action in the $SL(2)$ sector,
then  in section 3 review the  $(S,J)$ circular
string solution, and,  finally, in section 4 present
the computation of the 1-loop string correction to its energy.
In Appendix A, which is a generalization of section 2,
we shall discuss the ``fast string''
 limit of the \adss sigma model
in a more  general case  of five non-vanishing spins.
In Appendix B we shall give  some details
 of  computation of   fermionic
characteristic frequencies used   in section 4.
In Appendix C we shall discuss the Landau-Lifshitz frequencies
for the $(S,J)$ solution.


\renewcommand{\theequation}{2.\arabic{equation}}
 \setcounter{equation}{0}

 \section{``Fast-string''
  limit of string action in $SL(2)$ sector}

The SYM operators from the
$SL(2)$ sector  form a closed set under renormalization to all
orders in perturbation theory in $\l$ \ci{beis1,bs,bgk}.
As in the $SU(2)$ sector  \ci{kru,krt}, here  it makes
sense to  compare the string theory
and a coherent-state  spin chain
Lagrangians at large spins to all orders in the coupling
$\tilde{\lambda}=\frac{\lambda}{J^2}$.
This comparison  was initiated at
 the leading order  in \cite{st}.
 Here using a non-conformal ``homogeneous'' gauge similar to the
 one used  in the $SU(2)$ case
 \ci{krt,kt} we will show how
 to  compute subleading terms in the expansion of the string
 action for a general
 $(S,J)$ closed string configuration
   with a non-zero
 winding $m$ along a big circle in $S^5$ (ref.\ci{st} considered the
 $m=0$ case).
 The two-loop dilatation operator in the $SL(2)$ sector
 is not  known explicitly (yet the  corresponding Bethe equations and the
associated
 1-d S-matrix were recently found  in \ci{s}),
 so a direct comparison of the $\l^2$ subleading term
 in the string action  found  below to the 2-loop term in the
 coherent state ``Landau-Lifshitz'' action on the spin
 chain side (along the same lines as in  the $SU(2)$ case
 \ci{krt}) is not possible at present.

   \bigskip

Let us first set up our notation
by recalling the
 bosonic part of the $AdS_{5}\times S^{5}$ string action
 (see \ci{ft2,ts1,art,st})
\begin{equation}
I=\sqrt{\lambda}\int d\tau\int^{2\pi}_0
\frac{d\sigma}{2\pi}\sqrt{-g}(L_{S}+L_{AdS}) \label{stringact}
\end{equation}
\begin{equation}\la{dva}
L_{S}=-\frac{1}{2}g^{ab}\partial_{a}X_{M}\partial_{b}X_{M}+\frac{1}{2}\Lambda(X_{M}X_{M}-1)
\end{equation}
\begin{equation}\la{po}
L_{AdS}=-\frac{1}{2}g^{ab}\eta^{PQ}\partial_{a}Y_{P}\partial_{b}Y_{Q}+
\frac{1}{2}\tilde{\Lambda}(\eta^{PQ}Y_{P}Y_{Q}+1)
\end{equation}
where  $X_{M}\ (M=1, \dots, 6$) and $Y_{P}\ (P=0,
\dots, 5)$ are the embedding coordinates of $\mathbf{R}^{6}$
and of $\mathbf{R}^{2,4}$ (the latter with the metric
$\eta_{PQ}=(-1,+1,+1,+1,+1,-1)$).
Let us define 3+3 complex coordinates:
\begin{equation}\la{py}
\Y_{0}\equiv Y_{5}+iY_{0}\ , \quad \Y_{1}\equiv Y_{1}+iY_{2}\
, \quad
\Y_{2}\equiv Y_{3}+iY_{4}\ , \ \ \ \ \Y_{r}^{*}\Y^{r}=-1\ ,
\end{equation}
\begin{equation}\la{pya}
\X_{1}\equiv X_{1}+iX_{2}\ , \quad \X_{2}\equiv X_{3}+iX_{4}\ , \quad
\X_{3}\equiv X_{5}+iX_{6}\ , \ \ \ \ \X_{i}^{*}\X_{i}=1\ .
\end{equation}
Then  the \adss metric is
$
ds^{2}=d\Y_{r}^{*}d\Y^{r}+d\X_{i}^{*}d\X_{i}
$
where  $r,s=0,1,2$ and $i,j,k=1,2,3$
($\Y^{r}=\eta^{rs}\Y_{s}$, with $\eta^{rs}=(-1,1,1)$).
In terms of  the usual global time, radial  and
angular coordinates one has
\bea
&&\Y_{0}=\cosh \rho \ e^{it}, \quad \Y_{1}=\sinh \rho\  \sin \theta
\;e^{i\phi_{1}}, \quad \Y_{2}=\sinh \rho\  \cos \theta
\;e^{i\phi_{2}} \label{cartcoord1}\\
&&
\X_{1}=\sin \gamma \cos \psi \;e^{i\varphi_{1}}, \quad \X_{2}=\sin
\gamma \sin \psi \;e^{i\varphi_{2}}, \quad \X_{3}=\cos \gamma
\;e^{ i\varphi_{3}}\ .  \label{cartcoord2}
\eea
Separating the common phases of $\Y_r$ and $\X_i$
\begin{equation}\la{uv}
\Y_{r}=e^{iy}V_{r}, \quad \quad \X_{i}=e^{i\alpha}U_{i}, \quad
\quad V_{r}^{*}V^{r}=-1, \quad \quad U_{i}^{*}U_{i}=1\ ,
\end{equation}
the \adss  metric becomes:
\begin{eqnarray}\la{ki}
ds^{2}=- (Dy)^2+(D\a)^2+D^{*}V_{r}^{*}DV^{r}+D^{*}U_{i}^{*}DU_{i}
\end{eqnarray}
where
\begin{equation}
  Dy=dy+B \ , \quad \quad
DV_{r}=dV_{r}-iBV_{r} \ ,\quad \quad B=iV_{r}^{*}dV^{r}, \label{riv}
\end{equation}
\begin{equation}
D\alpha=d\alpha+C\ , \quad \quad
DU_{i}=dU_{i}-iCU_{i}\ , \quad \quad  C=-iU_{i}^{*}dU_{i}\  .    \label{deriv}
\end{equation}
In  what  follows we shall consider the string configurations
in  the $(S,J)$ sector where  the strings
may wound around  and move along a big circle in $S^5$
and may also move in $(1,2)$ plane in $AdS_5$, i.e.
\begin{equation}\la{yy}
\X\equiv \X_{1}=e^{i\alpha}, \qquad \qquad \X_{2}=\X_{3}=0
\end{equation}
\begin{equation}\la{ys}
\Y_{0}=e^{iy}V_{0}=\cosh \rho \;e^{it},
 \qquad \Y_{1}= e^{iy}V_{1} = \sinh \rho
\;e^{i\phi_{1}}, \qquad \Y_{2}=0\ .
\end{equation}
 We may fix  the obvious
$U(1)$ symmetry of $\Y_r$: \ $ y\rightarrow y-\chi , \ \  \quad
V_{r}\rightarrow e^{i\chi}V_{r}$
(from now on we will assume that $r,s=0,1$)
by the following choice of $y$
\begin{equation}
y=t \ ,  \label{fixy}
\end{equation}
so that  $
V_{0}=\cosh \rho, \ V_{1}=\sinh \rho \;e^{i(\phi_{1}-t)}$.
Then the  Lagrangian in \rf{stringact}
reduces to
\begin{equation}
L=-\frac{1}{2}\sqrt{-g}\ g^{ab}\left(-D_{a}tD_{b}t+\partial_{a}\alpha
\partial_{b}\alpha+D^*_{a}V_{r}^{*}D_{b}V^{r}\right)\ .
\label{Polyakov}
\end{equation}
As in \ci{krt} the  idea is  to assume that the string
is moving fast and gauge fix $t$ and the momentum conjugate to
the ``fast'' coordinate $\a$,
getting an effective  action for the ``slow'' transverse
coordinates $V_r$ only.

The conserved charges corresponding to
the translations in $t$,
rotations
 in $AdS_5$ or $\Y_1 \to e^{i \beta} \Y_1$
  and  translations
  in $\a$
\begin{equation}
E \equiv\sqrt{\lambda}\;\;\mathcal{E}, \qquad S\equiv
\sqrt{\lambda}\;\mathcal{S},\qquad J
\equiv\sqrt{\lambda}\;\mathcal{J}
\end{equation}
are given by
\begin{equation}\la{eee}
\mathcal{E}-\mathcal{S}=\int_{0}^{2\pi}\frac{d\sigma}{2\pi}\ p_{t},
\quad \quad   p_{t}=-\sqrt{-g}g^{0a}D_{a}t\ ,
\ee
\begin{equation}
\mathcal{S}=-\int_{0}^{2\pi}\frac{d\sigma}{2\pi}\sqrt{-g}g^{0a}
\big[V_{1}V_{1}^{*}D_{a}t
- (\frac{i}{2}V_{1}^{*}D_{a}V_{1}+c.c.)\big]
\label{spin}
\end{equation}
\be
\mathcal{J}=\int_{0}^{2\pi}\frac{d\sigma}{2\pi}\ p_{\alpha}\ , \ \
\ \ \ \ \ \ \ \ \
p_{\alpha}=-\sqrt{-g}g^{0a}\ \partial_{a}\alpha\  .
\label{charge}
\end{equation}
Following  \ci{kt}, we may   replace $\a$ by the dual coordinate
$\tilde{\alpha}$
\begin{equation}
\sqrt{-g}g^{ab}\partial_{b}\alpha=-
\epsilon^{ab}\partial_{b}\tilde{\alpha}\ .
\label{dual}
\end{equation}
Solving for   the 2-d metric $g_{ab}$ in \rf{Polyakov}
we get  the Nambu-type  Lagrangian
\begin{equation}\la{hh}
L=-\sqrt{h}\ , \ \ \ \ \ \ \ \ \ \  h=|\det h_{ab}|\ ,
\end{equation}
\begin{equation}
  h_{ab}=-D_{a}t
D_{b}t+\partial_{a}\tilde{\alpha}
\partial_{b} \tilde{\alpha} +D^*_{(a}V_{r}^{*}D_{b)}V^{r}
\label{h}
\end{equation}
To fix the world-sheet reparametrization freedom
 we choose,  as in \ci{krt,kt},  the following   two
gauge conditions
\begin{equation}\la{gau}
(i) \ t=\tau\ ,  \ \ \ \ \ \ \
(ii) \ \tilde{\alpha}=\mathcal{J} \sigma
\ , \ \ {i.e.} \ \ p_{\alpha}=\mathcal{J}=const \ .
\end{equation}
The first ensures that the space-time and the
world-sheet energies are
the same;  the second implies that $J$ is distributed
homogeneously along the string coordinate $\s$  (which
is the case on the spin chain side).
Then the string action becomes
\begin{equation}\la{ko}
I=J\int d\tau \int_{0}^{2\pi} \frac{d\sigma}{2\pi}\tilde{L}
\ , \ \ \ \ \ \ \ \
\tilde{L}=-  \J^{-1}  {\sqrt{h}} \ ,
\end{equation}
with ($ B_a = i V^*_r \del_a V^r $)
\begin{equation}\la{pi}
h=(\mathcal{J}^2-B_{1}^{2}+|D_{1}V_{r}|^2)[|D_{0}V_{r}|^2-
(1+B_{0})^2]-
[\frac{1}{2}(D_{0}V_{r}^{*}D_{1}V^{r}+c.c.)-B_{1}(1+B_{0})]^2
. \end{equation}
The next step \cite{kru,st,krt}
is to expand in large $\J$ assuming that
higher powers of time derivatives of $V_r$ are suppressed.
To define an  expansion in
$\frac{1}{\mathcal{J}^2}= {\l \ov J^2}
\equiv \tilde{\lambda}$ it is useful to
 rescale  $\tau$ so that
 the leading order term  does
not contain $\tilde{\lambda}$:\
$
\tau \rightarrow \mathcal{J}^2\tau\ , \
\partial_{0}\rightarrow \frac{1}{\mathcal{J}^2}\partial_{0} .$
Then
\begin{eqnarray}
\tilde{L}&=&-\mathcal{J}^2-B_{0}-\frac{1}{2}|D_{1}V_{r}|^2\label{LL1}\\
&+&\frac{1}{8\mathcal{J}^2}\left[|D_{1}V_{r}|^4+4|D_{0}
V_{r}|^2-4B_{0}|D_{1}V_{r}|^2+4B_{1}(D^*_{0}V_{r}^{*}D_{1}V^{r}+c.c.)\right]
+O({1\ov \mathcal{J}^4}) \nonumber
\end{eqnarray}
Finally, we may  eliminate the time derivative terms
by field redefinitions \ci{krt} or, to leading order, simply
 using the
``Landau-Lifshitz''  equation \ci{st} following from
\rf{LL1}\foot{To obtain this equation one should vary \rf{LL1}
taking  into account the constraint $V^s V^*_s =-1$ (which may be
imposed using, e.g., a Lagrange multiplier).}
\begin{equation}
i\partial_{0}V^{r}-\frac{1}{2}\partial_{1}^{2}V^{r}-\partial_{1}
V^{r}\partial_{1}V^{s}V_{s}^{*}-\frac{1}{2}\partial_{1}
(\partial_{1}V^{s}V_{s}^{*}) \ V^{r} \ + O({1\ov \mathcal{J}^2}) =0
\ . \label{motion1}
\end{equation}
We get
\begin{eqnarray}
\tilde{L}&=&-\mathcal{J}^2- i V^*_r \del_0 V^{r}
-\frac{1}{2}|D_{1}V_{r}|^2\label{LL2}\\
&+&\frac{1}{8\mathcal{J}^2}\left[4|D_{1}V_{r}|^4
+|D_{1}^{2}V^{r}|^2-
2( V^*_s \del_1 V_s)^{2}|D_{1}V_{r}|^2-2( V^*_s \del_1 V_s  D_{1}V^{r}
D_{1}^{2}V_{r}^{*}+c.c.)\right]\ .
\nonumber
\end{eqnarray}
As was shown in \ci{st},
the first non-trivial (order $\J^0$)
term here
matches the coherent-state  effective action
following from  the  1-loop dilatation operator
in the $SL(2)$ sector  \ci{bs,bgk}
(the Hamiltonian of the $SL(2)$ Heisenberg ferromagnetic).

Rescaling back $\tau\rightarrow \frac{1}{\mathcal{J}^2}\tau$,
we find that the corresponding
2d energy or Hamiltonian is
 (in view  of our gauge choice $t=\tau$,
is the same as the $E-S$  in (\ref{charge}))
\begin{eqnarray}
E-S=J&+&\frac{\lambda}{2J}\int_{0}^{2\pi}
\frac{d\sigma}{2\pi}
\ |D_{1}V_{r}|^2
\nonumber \\
&-&\frac{\lambda^{2}}{8J^{3}}\int_{0}^{2\pi}
\frac{d\sigma}{2\pi}\bigg[4|D_{1}V_{r}|^4+|D_{1}^{2}V_{r}|^2
 - 2 (V^*_s \del_1 V^s)^2    |D_{1}V_{r}|^2 \label{2denergy}\\
 && \ \ \ \ \ \ \ \ \ \ \ \ \ \  \ \ \ \  - 2(V^*_s \del_1 V^s  D_{1}V^{r}D_{1}^{2}V_{r}^{*} +c.c.)
 \bigg]
+ O( \frac{\lambda^{3}}{J^{5}}) \ .   \nonumber
\end{eqnarray}
 The expression for
the spin $\mathcal{S}$ in (\ref{spin})
expanded in large is
$\mathcal{J}$
\begin{eqnarray}
\mathcal{S}&=&\mathcal{J}\int_{0}^{2\pi}
\frac{d\sigma}{2\pi}V_{1}V_{1}^{*}\label{spin1} \\
&+&\frac{1}{2\mathcal{J}}\int_{0}^{2\pi}
\frac{d\sigma}{2\pi}\left[V_{1}V_{1}^{*}|D_{1}V_{r}|^2
-B_{1}(iV_{1}^{*}D_{1}V_{1}+c.c.)-
(iV_{1}^{*}D_{0}V_{1}+c.c.)\right]+O({1\ov \mathcal{J}^3})
\ , \nonumber
\end{eqnarray}
where the time derivatives should be again eliminated using
\rf{motion1}.

In the above discussion $\a$ was assumed to be a general
function of $\tau$ and $\sigma$, which, due to
 the periodicity
condition, should be subject to
\begin{equation}\la{win}
\alpha(\sigma+2\pi,\tau)=\alpha(\sigma,\tau)+2\pi m\ , \ \ \
{i.e.}\ \ \ \ \
 \int_{0}^{2\pi} \frac{d\sigma}{2\pi} \partial_{1}\alpha=m\ ,
\end{equation}
where $m$ is an integer
 winding number along the $S^5$ circle.
 A possible  winding  along the $AdS_5$ circle is
 (by  our  choice of $y=t=\tau$)  absorbed
 into $V_1$.
 Using  \rf{gau},(\ref{dual}) we find
$
m=-\mathcal{J}\int_{0}^{2\pi}\frac{d\sigma}{2\pi}\sqrt{-g}\ g^{01}
$, and after
plugging in the expression for the 2-d
metric $g_{ab}=h_{ab}$ in \rf{h}
we conclude that
\begin{equation}
m=-\mathcal{J}\int_{0}^{2\pi}\frac{d\sigma}{2\pi}
\frac{h_{01}}{\sqrt{h}}
\ ,  \ \ \ \ \ \ \ \ \
h_{01}=-B_{1}(1+B_{0})+\frac{1}{2}(D^*_{0}V_{r}^{*}D_{1}V^{r}
+c.c.)\ .
\end{equation}
This  is an additional constraint
which should be  imposed on any
particular solution of equations following from  (\ref{LL2}).
Expanding  this condition for large $\mathcal{J}$
and  eliminating time derivatives using \rf{motion1}
we get
\begin{equation}
m=\int_{0}^{2\pi}\frac{d\sigma}{2\pi}\
\bigg( i V^*_r \del_1 V^r
-\frac{1}{2\mathcal{J}^2}
\big[  i V^*_s \del_1 V^s |D_{1}V_{r}|^2
+\frac{1}{2}(iD_{1}V^{r}D_{1}^{2}V_{r}^{*}+c.c.)\big]\bigg)
+  O(\frac{1}{\mathcal{J}^4}) \ .
\label{constraint}
\end{equation}
We conclude that while the $S^5$ winding number $m$
does not enter  the effective Lagrangian,
it appears in the constraint on its solutions.
This is in agreement with the spin-chain side considerations
 where $m$
enters a  constraint on Bethe roots
but not the algebraic Bethe equations \cite{kz}
or the coherent-state effective action \ci{st}.


A  similar procedure of taking  a ``fast-string'' limit
of the string action   can
be applied  in a more general case of  $5$
non-vanishing spins
$(S_{1},S_{2},J_{1},J_{2},J_{3})$; we shall
discuss it  in
 Appendix A.\foot{Even though the corresponding set of SYM operators
 is, in general, not closed under renormalisation beyond one loop,
 it was argued in  \cite{minah}  that
non-closed sectors (like $SU(3)$ and
$SO(6)$)
 may be viewed as closed
in the thermodynamic limit even at higher loops.
 It then makes sense to compare the corresponding anomalous
 dimensions  or semiclassical
 effective actions they are described by
 to a limit of  string theory
 action  even beyond leading order in
the effective coupling  $\tilde{\lambda}$.}

\renewcommand{\theequation}{3.\arabic{equation}}
 \setcounter{equation}{0}

\section{Circular string solution in
$AdS_3 \times S^1 \subset AdS_5
\times  S^5 $}

In this section we shall review a remarkably  simple
circular  string  solution of \rf{stringact} found in  \cite{art}
 representing   a  particular
 semiclassical  $(S,J)$ state  in the  $SL(2)$ sector.
The string is positioned  in a plane in $AdS_3$ and is also
wrapped on $S^1$ in $AdS_5$.
 As we shall explain below, this configuration
  (its expansion in
large $\J$) can  be  obtained
also as a solution of the
``Landau-Lifshitz'' equations \rf{motion1}.

It is useful first to discuss its  flat-space analog
by considering a closed string  moving in
${\R}^{1,2}\times S^{1}$  instead of $AdS_3 \times S^1$
 (then  $ds^2 = - dt^2 + d\rho^2 + \rho^2 d \phi^2
  + R^2 d \a^2
$, \ $R$ is a  radius of $S^1$).\foot{Similar solution exists also
in the  twisted product (``Melvin'') case of
$ds^2 = - dt^2 + d \rho^2 + \rho^2 ( d \phi + q d \a)^2
 + R^2 d \a^2$.}
Let us use conformal gauge and consider a rigid circular string
lying in $\R^2$ (with coordinates $Y_1,Y_2$)
and rotating along itself while  being also  wound
along $S^1$  and
having non-zero $S^1$ momentum:
\begin{equation}
t=\kappa \tau\ ,\ \ \ \
 \quad \Y=Y_{1}+iY_{2}= \rho \;e^{i\phi} \ , \ \ \ \ \ \ \ \
 \phi = \varpi\tau+k\sigma \ ,
\qquad \a=w\tau+m\sigma\  ,
\end{equation}
The free string equations are solved if  the string
 radius is
constant,  $\rho=$const, and $\varpi=-k >0$ (another solution has
$\varpi=k$). The   polar angle in $\R^2$ is thus $\phi= |k|(\tau -
\sigma)$, i.e. $|k|$ may be interpreted as an  integer ``winding''
in $\R^2$ (we shall assume that $\kappa, m$ and $w$ are positive
and $k$ is negative). $m$ is an integer winding number in $S^1$
and the  linear momentum along $S^{1}$ is also quantized. This
configuration is thus a closed string wound along $S^1$ and having
a left (or right) moving fluctuation along it (at fixed time the
profile of the string is a spiral drawn on the $(\phi,\a)$
torus). The corresponding string state  should thus  be  1/2
 BPS in the  superstring theory.
Indeed, the  conformal gauge constraints imply
\begin{equation} \la{hoh}
   \kappa^{2}=2 \rho^2  k^2+ R^2 (m^2+w^2)\ , \ \ \ \ \ \
-\rho^2 k^2+ R^2 m w=0\ , \ \ \ \  {i.e.} \ \ \
\kappa = R( m + w) \ ,
\end{equation}
 and  the
   string energy $E  $, the spin $S$ in $\R^2$ and the
 $S^1$ momentum $J$  are  given by
\begin{equation}
(E,S,J)= { 1 \ov \a'} ( \E,\S,\J) \ , \ \ \ \ \ \
\E=R( m +  w) \ , \quad \quad  \S= \rho^2 |k| =
R^2 \frac{m w}{|k|}, \quad \quad \J= R^2  w\  \ ,
\end{equation}
where ${ 1 \ov 2\pi \a'} $ is the string tension.
The constraints \rf{hoh} expressed  in terms of
$E$,$S$,$J$,$k$ are
\begin{equation} \la{hih}
\E = R^{-1}  \J +  R |k| { \S \ov \J}  \ , \ \ \  \ \ \ \ \ \ \ \ \
k \S + m \J =0\ .
\end{equation}
If we now replace ${\R^{1,2}}$ with the $AdS_{3}$ subspace
of $AdS_{5}$ and consider a similar  string  configuration
in $AdS_3 \times S^1$ we obtain the circular solution
 \cite{art}  in
$AdS_{5}\times S^{5}$. It, however,
 will no longer be a BPS state, and  will not reduce
 to the above flat-space solution in the limit  of large radius
 of $AdS_{5}\times S^{5}$. Indeed, here
 the radii of $AdS_5$ and $S^5$ are the
same (with  ${ R^2 \ov \a'}= \sqrt \l$), so that
taking $R$  large to approximate $AdS_3$ by
$\R^{1,2}$ would also make the radius of $ S^1$
large, which would correspond to
sending the energy of the flat-space solution
\rf{hih} to infinity.

Let us now  review this circular solution of the
\adss equations following from (\ref{stringact})
by using the conformal gauge.
It is characterised  by having
constant Lagrange multiplies $\Lambda,\tilde{\Lambda}$
and  constant induced metric. Setting $\Y_2=0, \ \X_2,\X_3=0$
(i.e. specialising to the $SL(2)$ sector \rf{yy},\rf{ys} or
to the motion in $AdS_3 \times S^1
\subset $\adss) one finds:\foot{The $AdS_5$ part of this  solution can also be expressed in
the Poincare coordinates  ($ds^2 = {dz^2 +  dx_m dx_m \ov z^2}$)
as follows: $x_{0}=\tan \kappa \tau$, $x_{1}=\tanh
\rho_{0} \;\frac{\cos(\varpi \tau+k\sigma)}{\cos \kappa\tau}$,
$x_{2}=\tanh \rho_{0} \;\frac{\sin(\varpi \tau+k\sigma)}{\cos
\kappa\tau}$, $x_{3}=0$, $z=\frac{1}{\cosh \rho_{0} \cos \kappa\tau}$.}
 \bea
 &&  \Y_{0}=\r_{0} \; e^{i\kappa \tau}\
 , \qquad \Y_{1}=\r_{1} \;e^{i\varpi \tau+ik\sigma}\ ,
  \qquad \X_1=e^{iw\tau+im\sigma}\ ,
  \label{sol}
 \eea
 \be \la{rad}
 \r_0 \equiv \cosh \rho_{0}\ , \quad\quad \ \ \ \
 \r_1\equiv \sinh \rho_{0}\ , \ \ \ \ \ \ \ \ \ \ \ \ \
  \r_0^2 - \r_1^2 =1 \ ,
 \ee
where $\rho_{0}$ is a  constant radius of the circular string,
$k$ and $m$ are the winding numbers in the  2-plane in $AdS_3$ and in
$S^1$, and  $\varpi $ (which is no longer
 equal to $k$ as in the
flat space)
and $w$ are rotation frequencies of the string along
itself.\foot{Note that this solution cannot be analytically
continued directly to a
physical solution in the $SU(2)$ sector as was done for a
folded string solution in \ci{bfst} since that would
introduce windings in the
transformed time ($\alpha \to t$); one will need also
to redefine the world-sheet coordinates to keep both $AdS_5$
time coordinates single-valued.}


 Since the angles
$\phi_1= \varpi \tau+k\sigma$ and $\a= w\tau+m\sigma$
in \rf{yy},\rf{ys} (which correspond to isometries of the metric
and thus enter the string Lagrangian only through their
derivatives)
are linear in the world-sheet  coordinates,
their derivatives are constant, and thus the induced
metric and also  all the coefficients in
the  fluctuation Lagrangian near this
 solution  (to all orders in the fluctuations)  are constant.
The equations of motion imply:
 \bea
 && \varpi^{2}=\kappa^2+   k^2, \qquad w^2=\nu^2 + m^2 , \qquad
 \nu^2= -\Lambda, \qquad \k^2=\tilde{\Lambda} \ . \label{pararel}
\eea
In terms of the  non-zero  charges
\begin{equation}
(E,S,J)=\sqrt{\lambda} (\E,\S,\J)\ , \qquad
\E =  \r^2_0\kappa\ , \qquad
\S= \r^2_1 \varpi  \ , \qquad  \J=  w\ ,
\end{equation}
the conformal gauge constraints are (we assume that
$\S,\J,m > 0 $  and $k< 0 $)
\be
2\kappa
\mathcal{E}-\kappa^2=2\sqrt{k^2+\kappa^2}\mathcal{S}
+\mathcal{J}^2+m^2\ ,
 \label{cgc}\ee
\be \la{kik}
 k\mathcal{S}+m\mathcal{J}=0\ ,
\ee
and \rf{rad} gives  also
\be \frac{\mathcal{E}}{\kappa}-\frac{\mathcal{S}}
{\sqrt{k^2+\kappa^2}}=1 \  . \la{rell}\ee
As implied by  these relations,   there
are only three independent parameters,
e.g., $\S,\J,k$  (which are
useful
for comparison with gauge theory),
or $\k,r_1, k$ (which are useful for finding fluctuation
 frequencies
discussed below in sect.4).
Eqs.  \rf{cgc},\rf{kik},\rf{rell}  imply that
in terms of $\k,k,\r_1$
\be\la{nuu}
\nu^2=\sqrt{\J^2 - m^2} = \sqrt{(\kappa^2-2k^2\r_{1}^{2})^2-
4k^2\r_{1}^{4}(k^2+\kappa^2)}\ , \ \ \ \ \ \  \ \ \
m^2 =  \ha \big( \kappa^2-2k^2\r_{1}^{2}  -  \nu^2 \big) \ .
 \ee
 In terms of $\J, \S$  and $k$
 we find that for  large $\J$  and $\S$  with
  fixed $u\equiv { \S\ov \J}$  and $k$ \foot{In this case
   the string
  motion is
 ``fast'' \ci{mik12} meaning that each point of the string moves
 at almost the speed of light and thus
   the induced metric degenerates.}
$ \nu=\sqrt{ \J^2 - k^2 u^2} = \J - \frac{k^2}{{2\J}} u^2 + ...$,
and
\be\la{ku}
 \k= \mj+\fr{k^2}{2\J}u (2+u)
 -\fr{k^4}{8 \J^3} u (4 +12 u+8u^2   +u^3 )+\cdots
  \ , \ \ \ \ \ \ \ \     u\equiv \frac{\S}{\J} \ .  \ee
\be \varpi= \sqrt{ \kappa^2 + k^2} = \J + { k^2 \ov 2 \J} (1
+u)^2+ ...\ , \ \ \ \ \ \r^2_1= {\S \ov \sqrt{ \k^2 + k^2} }
= u-\fr{k^2}{2\J^2} u (1+u)^2 + ... \ , \ee
and the energy $E=\sqrt\l\ \E(\S,\J,k)=E(S,J,k;\l)$
becomes (cf. \rf{cla})
\begin{eqnarray}
E&=&J (1+u) \bigg[1+\frac{\lambda k^2}{2J^2} u
-  \frac{\lambda^2 k^4 }{8J^4} u  (1 + 3 u + u^2)
 +O({\lambda^3 \ov J^6})\bigg]\ ,  \ \ \ \ \ \
  u\equiv \frac{S}{J} \ .    \label{tw}
\end{eqnarray}
Given that  the string energy looks like
 a regular expansion in  $\lambda$, it
was suggested in \cite{art} (by analogy
with previous results for string  solutions
in $SU(2)$  and $SU(3)$ sectors  \ci{ft2,afrt,bmsz,mez,kri})
  and later verified in \cite{kz}
 that the  leading  order $\l$ term in \rf{tw}
can  be reproduced as a one-loop anomalous dimension of a
 SYM  operator from the $SL(2)$ sector.

\bigskip

Finally, let us discuss
how  the above winding circular string  appears
(at leading order in $\mathcal{J}$)  as a solution
of the ``Landau-Lifshitz''
equation \rf{motion1}.\foot{Circular and folded
string solutions  of the
Landau-Lifshitz equations in the $(S,J)$ sector were also
discussed   in \ci{ryang}.
In the case of the circular solution, it is important
not to do field redefinitions that  may  contradict
the required single-valuedness condition for the $AdS_5$ time
  $t(\sigma+2\pi,\tau)=t(\sigma,\tau)$.
 }
Concentrating first on the $S^5$ part of the configuration
we note that
 the world-sheet metric
$h_{ab}$ has, to  leading order, the following components $
h_{00}=-1, \ h_{01}=-B_{1}, \ h_{11}=\mathcal{J}^2, \
h=\mathcal{J}^2+B_{1}^2$.
Then
(\ref{dual}) and \rf{gau} imply
(after the rescaling of $\tau$ so that $t= \J \tau $)
that
$\alpha=\mathcal{J}\tau+m\sigma, $
 which  is the same as the phase  of  $\X_1$
 in  \rf{sol}.
Turning to  the $AdS_5$ part, the Lagrangian  corresponding to the
leading-order term in (\ref{LL2}) is (after rescaling back
$\tau\rightarrow \frac{1}{\mathcal{J}^2}\tau$)
\begin{equation}
L =\dot{\eta}\sinh^{2}\rho -\frac{1}{2
\J^2}\left(\rho'^2+\eta'^2\sinh^{2}\rho\ \cosh^{2}\rho\right)\ ,
\label{LL4}
\end{equation}
where
we used \rf{ys} and introduced $\eta= \phi_1 -t$ (dot and prime
are $\del_\tau$ and $\del_\sigma$).\foot{Note that after the field
redefinition $\eta\rightarrow -2\eta$ this Lagrangian is the same
as (2.31) of \cite{st}.} The corresponding equations of motion
have solution with $\rho=\rho_{0}$=const provided
 $\eta''=0$.
The solution for $\eta$ is then found to be
$ \eta=q\tau +   k\sigma \ ,$  \ \
$ q= \frac{k^2}{2\J^2} (1+2 \r_{1}^2),$
implying that
$\phi_{1}=\mathcal{J}\tau+k\sigma$, which indeed
is the leading order form of $\phi_1$ in \rf{sol}.
Using (\ref{2denergy}) one can compute the space-time energy
which  reproduces the order $\l$  term in
(\ref{tw}).
Also, the ``winding'' constraint (\ref{constraint}) leads,
 to  the leading
order, to  the same condition that follows in general
 from the  conformal gauge
constraint $m\mathcal{J}+k\mathcal{S}=0.$

\renewcommand{\theequation}{4.\arabic{equation}}
 \setcounter{equation}{0}

\section{One-loop correction to energy of circular solution }

As was already mentioned, for all ``homogeneous''  circular
solutions \ci{art} like the one of the previous section  \rf{sol}
 the fluctuation Lagrangian has constant
coefficients (to all orders in fluctuation fields)
\ci{art,ts1}, a property shared with the BMN case
where one expands near a point-like geodesic \ci{ft1,par,callan}.
 That  means, in particular,
  that the spectrum of quadratic fluctuations
  can be found  explicitly (and one can,
  in principle,  also compute subleading
  corrections to their energies as in  the BMN case in
  \ci{callan}).
   The
general procedure of how to do that for a  generic
quadratic 2-d
Lagrangian of the type  ($p=1,...,N$, and
$K,W,M$ are constant matrices)
\be \la{lkt} L_2 = \dot  x_p^2
- x'^2_p  +  K_{pq} x_p \dot x_q +  W_{pq} x_p x'_q
+ M_{pq} x_p  x_q \ , \ee
was discussed  in \ci{blau,ft2,ft3}.\foot{Such Lagrangian is
readily obtained for the bosonic fluctuations;
similar first order or second order
 one is found for the fermionic fields.}
Assuming that the fields are periodic in $\s$, one
 sets \be
  x^p (\tau,\sigma) = \sum^\infty_{n=-\infty}
x^p_n (\tau) e^{i n \s}\ , \ \ \ \ \ \ \
x^p_n (\tau)= \sum^{2N}_{I=1} A^p_{I,n} e^{i \om_{I,n}\tau} \ , \ee
where $I$ labels $2N$   ``phase-space'' directions.
Then the field equations lead to
a  system of linear homogeneous equations for $A^p_{I,n}$, i.e.
$ F_{pq} A^q_{I,n}=0$, where  $ F_{pq}$ depends on
$n, \om_{I,n}$ and the coefficients in the Lagrangian \rf{lkt}.
It has solutions provided  det$F_{pq}=0$, which gives an
order $2N$  polynomial equation for the characteristic
 frequencies.
 As follows from the reality conditions, which translate into
 $F^{T}( \om_{I,n},n)=F(- \om_{I,n},-n)$,
 the zero-mode ($n=0$) frequencies  come
 in pairs, i.e. $\om_{I,0}= \{ \pm \om_{p,0}\}$
 ($p=1,...,N$). The
 non-zero frequencies can be labelled so that they
satisfy $\om_{I,n}= - \om_{I,-n}$,
 i.e. can also be paired
  (and associated to  creation
 and annihilation operators upon quantization).
 Then the
 one-loop correction to the 2d ground state energy
 is given by the following sum of the non-trivial half of the
  characteristic frequencies \ci{blau},
\bea
 E_{2 \rm d} =  \fr{1}{2}\sum_{p=1}^{N} \hat  \w_{p,0}
   +\fr{1}{2}\sum_{n=1}^\infty\;\sum_{I=1}^{2N} \hat \w_{I,n}\ .
 \label{e}
 \eea
 Here
\be
\hat  \w_{p,0}
 =   sign(C_{p}^{B}) \w_{p,0}\ , \ \ \ \ \ \ \ \
\hat \w_{I,n} =  sign(C_{I,B}^{(n)})  \w_{I,n} \ , \ee
 \begin{equation}
C_{p}^{B}=\frac{1}{2m_{11}(\w_{p,0})\w_{p,0} \prod_{q\neq
p}(\w_{p,0}^{2}-\w_{q,0}^{2})},\quad
C_{I,B}^{(n)}=\frac{1}{m_{11}(\w_{I,n})\prod _{J\neq
I}(\w_{I,n}-\w_{J,n})}\ ,  \label{signs}
\end{equation}
where $m_{11}$ is a minor of $F$, i.e. the determinant of the
matrix obtained from $F$ by removing the first row and first
column.\foot{
 For example, for a single real ($N=1$)
  2d field with mass $\mu$
 the characteristic equation is
 $\om^2 = n^2 + \mu^2$, so that
 $E_{2 \rm d} = \fr{1}{2} \sum_{n=-\infty}^\infty \sqrt{ n^2 + \mu^2}
 = \fr{1}{2} \mu + \sum_{n=1}^\infty \sqrt{ n^2 + \mu^2}.$ To see that
 this is the same as \rf{e} we note that in the case of a free
 massive field $sign(C_{p}^{B}) \w_{p,0}=|\w_{p,0}|$,
  $sign(C_{I,B}^{(n)})
 \w_{I,n}=|\w_{I,n}|$. (Note that this relation is not true
in general -- it is true only when half of the frequencies
are positive and half are  negative for $n>0$.)
 This can be seen by writing the matrix
 $F$ for a system of two first-order equations in time:
 $F_{11}=n^2+\mu^2$, $F_{12}=i \w_{I,n}$, $F_{21}=-i \w_{I,n}$,
 $F_{22}=1.$
Then  \rf{e} gives again $\fr{1}{2}
 \mu + \sum_{n=1}^\infty \sqrt{ n^2 + \mu^2}$
 since for a single massive field with mass $\mu$:
  $|\w_{1,n}|= |\w_{2,n}| = \sqrt{ n^2 + \mu^2}$.}

  In  the  case of the circular string solution
  the number  of  transverse fluctuations is $N=8$
  and since we are to start with the superstring action
  \ci{MT} we also have the fermionic modes (see \ci{ft1,ft2,ft3}
  for details).\foot{As we shall see below, here
   both the bosonic and
  the fermionic 2d fields are
   periodic in $\s$ so that the sum over $n$ is
  over the integers.}
  Since in \rf{sol} one has  $t=\k \t$,  the 1-loop
correction to the space-time energy
$E_1$ is then given by
 \begin{eqnarray}
E_1=\fr1{\k}E_{2{\rm d}}&=&\fr{1}{2\k}\bigg[ \sum_{p=1}^{8}\left(
     \hat w_{p,0}^B -\hat \w_{p,0}^{F}
    \right)
   +   \sum_{n=1}^\infty\;\sum_{I=1}^{16} \bigg(
    \hat\w_{I,n}^B- \hat\w_{I,n}^{F}\bigg) \bigg] \ .
 \label{e1}
 \end{eqnarray}
Finally, one would like   to  expand the parameters
in $E_1$ at large $\J$
and compare  the leading asymptotics of $E_1$ with the subleading
term in the SYM anomalous dimension \rf{sym}.

The first step is thus
 to  find, as in \ci{ft3}, the bosonic and fermionic
Lagrangians  for small fluctuations near the circular solution
\rf{sol} and then to compute the corresponding characteristic
frequencies.
This will be done in the next two subsections 4.1 and 4.2, where
we shall also  verify the stability of the solution,
 i.e. the reality of the characteristic frequencies
 and thus  of $E_1$ for  large enough  angular momenta.
 In section  4.3   we shall  check
  the  UV finiteness of
 (\ref{e1}) which is
  a consequence of the conformal invariance of the
 \adss superstring theory \ci{MT,ft2,ft3}. We shall
 also note a possibility of evaluating the asymptotics of
 $E_1$ by  interchanging summation over modes
 with  large spin expansion.
  Since the analytic evaluation of the
  sums involved  does not  appear to be possible  we shall then
  follow   ref.\ci{fpt}  and  compute  the leading asymptotics
  of $E_1$ at large $J$ and fixed $S/J$  using a numerical method.

\subsection{Bosonic  frequencies}

Below we shall review the
 derivation of the bosonic characteristic
 frequencies for the solution \rf{sol}
in  the conformal gauge  following   \ci{art}.
Let us consider the $S^5$-directions first. In general, starting
with \rf{dva}--\rf{pya} and expanding $ \X_i \to \X_i + \tilde{\X}_i$
one finds the quadratic fluctuation Lagrangian
in terms of the 3  complex fields
 \bea
 \tilde{L}_{S}=-\fr12\;\pa_a\tilde{\X}_i\pa^a\tilde{\X}_i^*
 +\fr12 {\Lambda}\;
  \tilde{\X}_i\tilde{\X}_i^*\ , \ \ \ \ \ \ \
 \sum_{i=1}^{3}(\X_i \tilde{\X}_i^*+\X_i^* \tilde{\X}_i)=0\ .
 \label{Xconstr}
 \eea
For the solution (\ref{sol}) $\tilde{\X}_{2}, \tilde{\X}_3$ are
decoupled  and represented by 4 real massive (with
mass $\nu$, see \rf{pararel})    2d fields, i.e. they
contribute  to the  energy \rf{e1} the term
 \bea
 {1 \ov 2 \k}  ( 4 \nu +
 2 \sum^\infty_{n=1}  \  4  \sqrt{n^2+\n^2} \ ) \ ,
\ \ \ \ \ \  \ \ \ \  \n^2  = \J^2 -m^2\ . \label{spheref}
 \eea
The Lagrangian for the fluctuations in the  $AdS_5$ directions is
 \bea
 \tilde{L}_{AdS}=-\fr12\;\pa_a\tilde{\Y}^r\pa^a\tilde{\Y}_r^*
 -\fr12 \tilde{\Lambda}\;
  \tilde{\Y}^r\tilde{\Y}_r^*
 \ , \ \ \ \ \ \ \ \ \
 \sum_{r=0}^2\;({\Y}^r\tilde{\Y}_r^*+{\Y}_r^*\tilde{\Y}^r)=0\ .
  \label{AdSconst}
 \eea
For the  solution (\ref{sol}) the mode  $\tilde{\Y}_2$ is
a decoupled massive complex field or two real  fields, i.e.
 it contributes to \rf{e1} as
 \bea
  {1 \ov 2 \k}  ( 2  \k +
 2 \sum^\infty_{n=1}  \  2  \sqrt{n^2+\k^2}\ ) \ . \label{adsf}
 \eea
Setting for  the remaining fluctuation fields
 \be
 \tilde{\X}_1=e^{iw\t+im\s}(g_1+if_1)\ , \ \  \
  \tilde{\Y}_0= e^{i\kappa \t}(G_0+iF_0)\ ,\ \  \
  \tilde{\Y}_1= e^{i\varpi\t+ik\s}(G_1+iF_1)\ ,
 \ee
we can solve the conditions in
 (\ref{Xconstr}) and (\ref{AdSconst}) as
 \bea
 g_1=0 \ ,\ \ \ \ \    \qquad G_0= {\r_1 \ov \r_0} G_1\ ,  \label{Gconsta}
 \eea
so that the Lagrangian for the 4 real fields
 $(f_1,F_0,F_1,G_1)$ becomes
 \bea
\tilde{L} =&& \fr12({\dot{f}_1}^2-{f'^2_1})
-\fr12 (
            \dot{F}_0^2-{F_0'}^2
                       )+2\fr{\r_1}{\r_0}\k F_0\dot{G}_1
                \nn\\
  && +\fr12\big(
            \dot{F}_1^2+\fr{1}{\r_0^2}\dot{G}_1^2-{F_1'}^2
            -\fr{1}{\r_0^2}{G_1'}^2
                       \big)-2\varpi F_1\dot{G}_1+2k F_1G_1'
 \ .  \label{ba}
 \eea
Note that $f_1$  couples
to $(F_0,F_1,G_1)$ through the conformal gauge conditions.\foot{The
 conformal gauge
constraints have the form
$ 2\r_1k^2G_1-\r_0\k\dot{F}_0+\r_1(\varpi\dot{F}_1+kF_1')
  +w\dot{f}_1+mf_1'=0  ,$  \ \ $
   -\r_0\k F_0'+2\r_1\varpi kG_1+\r_1(\varpi F_1'+k\dot{F}_1)
 +wf_1'+m\dot{f}_1=0 . $
}
 After an  appropriate
linear field redefinitions two of these four modes become
 massless and decouple,
and their contribution to the 1-loop effective action
(which does  not depend on the parameters of the background)
 gets cancelled by the two conformal ghost contributions.
  The upshot, as in \ci{ft3}, is that
 one can simply ignore the constraints on fluctuations
 implied by the conformal gauge conditions and omit two massless
 longitudinal modes.\foot{The same conclusion is found by
 starting with the Nambu action and
 using  static gauge condition on the fluctuation fields.}
Then the 4 characteristic frequencies corresponding to the two
remaining real 2d fields  can be found from  the following matrix
$F$
\begin{eqnarray}
\left(%
\begin{array}{cc}
  \frac{\w^2-n^2}{1+\r_{1}^{2}}+\frac{4k^2 \kappa^2\r_{1}^{2}}{A} & -\frac{2ik\r_{1}}{A}[\omega(k\varpi-m w)+n(k m-w \varpi)] \\
  \frac{2ik\r_{1}}{A}[\omega(k\varpi-m w)+n(k m-w \varpi)] & (\w^2-n^2)\left(1-\frac{\varpi^2-m^2}{A}\right) \\
\end{array}%
\right)
\end{eqnarray}
where $A=\frac{1}{2}(\sqrt{\kappa^4-4\kappa^2 k^2 \r_{1}^2
(1+\r_{1}^2)}-\kappa^2).$ The $\det F=0$ condition gives the
quartic
 equation (we use $\k,\r_1,k$ as independent parameters)
 \ci{art}
 \bea
 (\w^2-n^2)^2+4\r_1^2\k^2\w^2-4(1+\r_1^2)(\sqrt{\kappa^2 + k^2 } \;
 \w -kn)^2
 =0 \ . \label{bosch}
 \eea
The 4 roots  $\w= \{\w_{I,n}\}$ ($I=1,2,3,4$)
 are the characteristic  frequencies
in the  discussion at the beginning of this section
(here  $ \sum^4_{I=1} \w_{I,n}=0$
since there is no $\w^3$ term in \rf{bosch}).
Note that since \rf{bosch} is invariant under
 $\w\to - \w, \ n \to - n$,  we have, as required, the pairing
 of $n \not=0$ frequencies:
  $\w_{I,n}= - \w_{I,-n}$.  The zero modes are given by
  (using \rf{pararel})
  \be\la{zz}
\w_{I,0}=\{0,0, \w_0, - \w_0 \} \ , \ \ \ \ \ \
\w_0 =2 \sqrt{\kappa^2+ (1+\r_1^2) k^2 }\ , \ee
so that the contribution of these  ``mixed''  modes
to the bosonic part of \rf{e1} is thus
 \bea
  {1 \ov 2 \k}  \left( \w_0     +
  \sum^\infty_{n=1} \sum_{I=1}^4  sign(C_{I,B}^{(n)})\w_{I,n}\  \right)   \ .  \label{sfq}
 \eea
The relevant part of the minor $m_{11}$ for computing the signs of
$C_{I,B}^{(n)},C_{p}^{B}$ is $m_{11}\sim(\omega^2-n^2)$, since
$\left(1-\frac{\varpi^2-m^2}{A}\right)$ is always positive.
  While  the 4   roots $\w_{I,n}$ of the quartic equation
  \rf{bosch}
    can be
written down explicitly,  the resulting sum in \rf{sf}
cannot be done  analytically for generic values of the parameters.
Below we shall discuss a  way to evaluate the
sum in the large spin limit we are interested in.

In order for the circular string configuration
 to be stable all bosonic frequencies
must be real, and this is not
a priori obvious  for the solutions of \rf{bosch}.
If one expands $\om_{I,n}$ at large  $\kappa$ or
 large $\mathcal{J}$
(with $n$ and $\r_1^2\approx u, k$  being fixed, cf. \rf{ku}),
 one finds     that the frequencies are
real \ci{art}\foot{They are also  always
real for large $n$.  Numerical analysis
shows that for any given $n$ there  always exists
  a large enough $\mathcal{J}$ for which all 4
   frequencies are real.
For  small values of $\mathcal{J}\sim 1$
and  some small $n$ the solutions of
(\ref{bosch}) become complex.
 The stability   of the circular $(S,J)$ solution \rf{sol}
at   large spin $\mathcal{J}$  is similar to the stability of
 the
3-spin $S^5$ solution $(J_1=J_2, J_3)$ which is also stable for
large enough $J_3$ \ci{ft3}.}
\begin{eqnarray}
\omega_{I=1,2;n}&=&\frac{n}{2\mathcal{J}}\left[ 2k(1+\r^2_1)\pm
\sqrt{n^2+4k^2\r^2_1 (1+\r^2_1)}\right]+
O({1\ov \mathcal{J}^3})\ , \la{ol} \\
\omega_{I=3,4;n}&=&\pm 2\mathcal{J}\pm
\frac{1}{2\mathcal{J}}[n^2\mp
2kn(1+\r^2_1)+2k^2(1+3\r^2_1+\r^4_1)]+O({1\ov
\mathcal{J}^3}) \ . \la{pol}
\end{eqnarray}
As a result, the expression for $E_1$ is real and its comparison
to gauge-theory anomalous dimension is unambiguous (in contrast to
the case of the circular solution from $SU(2)$ sector discussed in
\ci{ft3,fpt} which is unstable for any $J_1=J_2$). Using the above
expressions of the frequencies at large $\mathcal{J}$ the
corresponding signs of $C_{I,B}^{(n)}$ are found to be:
$sign(C_{1,B}^{(n)})=sign(C_{3,B}^{(n)})=+1$ and
$sign(C_{2,B}^{(n)})=sign(C_{4,B}^{(n)})=-1.$ They are needed
later to compute the energy $E_{1}$ in the same limit.

Let us mention that out of all  the fluctuation  frequencies only
the two  in \rf{ol} scale as ${a \ov \J}$ at large $\J$, while all
other (including the fermionic frequencies discussed below) scale
as $ \J + { b \ov \J}$. The two  bosonic  frequencies of the
former  type have been reproduced from the
 Landau-Lifshitz
action in \rf{freq}: they correspond to deformations along the
``intrinsic'' $SL(2)$ sector directions $\rho$ and $\phi_1$ in
\rf{ys}. These two are then the ones  that should be ``seen'' also
on the $SL(2)$ spin chain side \ci{st,kz} (to reproduce  other
frequencies  one is to  embed the solution
into a larger  sigma model or  spin chain sector).

\subsection{Fermionic  frequencies}

 The  quadratic part of the \adss
superstring  Lagrangian evaluated on a bosonic solution
has  a simple form (see \ci{MT,ft1,ft2,ft3} for details)
 \bea
 L_F=i\left(\eta^{ab}\d^{IJ}-\e^{ab}s^{IJ}\right)\;
 \bar{\th}^I\rho_aD_b\,\th^J\;\;,\;\;\ \ \ \rho_a\equiv \G_A
 e_a^A\ , \;\;\;\; e_a^A\equiv E_\m^A(\x)\pa_a \x^\m\ ,
 \eea
where $I,J=1,2,\;s^{IJ}=\mbox{diag}(1,-1),$ \ $ \rho_a$ are
projections of the ten-dimensional Dirac matrices and $\x^\m$ are
 the coordinates of the $AdS_5$  space for $\m=0,1,2,3,4$ and
 the coordinates of $S^5$
for $\m=5,6,7,8,9$. The covariant derivative is given by
 \bea
 && D_a\th^I=\left(\d^{IJ}{\rm D}_a-\fr{i}2\e^{IJ}\G_*\rho_a\right)\th^J
 \;,\ \ \ \ \quad  \G_*\equiv i\G_{01234}\;,\;\ \G_*^2=1 \ ,
\eea
where ${\rm D}_a=\pa_a+\fr14\w_a^{AB}\G_{AB}, \ \  \w_a^{AB}\equiv
 \pa_a  \x^\mu  \w_\mu^{AB}$.
Fixing the $\k$-symmetry by the same condition as in \ci{ft3} \
 $\th^1=\th^2=\th $ \
one gets
 \bea
  L_F=-2i\bar{\th}D_F\th\;,\quad\quad  D_F=-\rho^a {\rm
  D}_a-\fr{i}2\e^{ab}\rho_a\G_*\rho_b\ .
 \eea
Labelling the coordinates as follows
(cf. \rf{cartcoord1},\rf{cartcoord2}):
 \bea
\m: && 0\;\;1\;\;2\;\;3\;\;\;4\;\;\;\;5\;\;\;6\;\;\;\;7\;\;\;8\;\;\;\;9\nn\\
\x^\m: &&
t\;\;\rho\;\;\th\;\;\phi_1\;\;\phi_2\;\;\g\;\;\varphi_1\;\;\psi\;\;
 \varphi_2\;\;\varphi_3
 \eea
one finds (using \rf{sol},\rf{rad})
that the non-trivial  components of the Lorentz connection
$\w_a^{AB}$ are
$
 \w_\t^{10}=-\k \r_1,  \;
 \w_\t^{13}=-\varpi \r_0, \;
 \w_\s^{13}=-k\r_0\ ,
$
and then
 \bea
  D_F
 &=&\left(\k\r_0\,\G_0+\varpi\r_1\G_3+ w \,\G_6\right)
     \left(\pa_\t-\fr12 \k \r_1 \G_{10}
       -\fr12\varpi\r_0\,\G_{13} \right)\nn\\
  &&-(k\r_1\,\G_3+m\,\G_6)\left(\pa_\s-\fr12k\r_0\G_{13}\right)
 +k\k \r_1\r_0\G_{124}\ .  \label{dfo}
 \eea
We observe that in contrast to the case of the $S^5$ circular
string
solution considered in \cite{ft3,fpt}
here $D_F$ does not depend on $\sigma$, and thus
one does not need to apply a  local rotation to eliminate
this $\s$  dependence (which in \cite{ft3,fpt} was
making the rotated
fermions antiperiodic in $\s$).
To simplify $D_F$ it  is useful  to do constant rotations
 in (36)-plane and (06)-plane (see Appendix B)
  which  do  not affect
 the periodicity  of the fermions,
 so the sum over the fermionic
   characteristic frequencies
 will go over integer $n$ as  in \rf{e1}.

 Since $\theta$ is a Majorana-Weyl 10d spinor,
 we end up with  the total of 16 characteristic frequencies
 (i.e. $I=1,...,16$)
 corresponding to 8 physical 2d fermionic modes.
 Their contribution to \rf{e1}
  expressed in terms of the
 three independent parameters $\kappa, \r_{1}^{2},k$ is
 found to be (see Appendix B for  details)
 \bea
 - {1 \ov 2 \k}  \bigg(  8 \td \w_0     +
  8 \sum^\infty_{n=1} \big[ \sqrt{  ( n + c)^2 + a^2}
   + \sqrt{ (n - c)^2 + a^2 }\ \big]
          \ \bigg)   \ ,  \label{sf}
 \eea
where
\be  \td \w_0 = \sqrt{ c^2 + a^2 } \ , \ee
\begin{equation}\la{cav}
a^2 = \frac{1}{2}(\kappa^2+\nu^2 ) \ , \qquad \quad
\nu^2= \sqrt{(\kappa^2-2k^2\r_{1}^{2})^2-
4k^2\r_{1}^{4}(\kappa^2 + k^2 )}\ ,
\end{equation}
\be \la{cavv}
 c   =\ha \kappa \big[ 1
+{2k^2(1+\r_{1}^2)\ov  \k^2 - \nu^2 }  \big]
  \sqrt\frac{\kappa^2-\nu^2 -2k^2r_{1}^{2}
 }{2(\kappa^2 + k^2)}\  .
\end{equation}
The  large $\mathcal{J}$ expansion of the
fermionic frequencies in  \rf{sf}  is
(the expansion of $\td \w_0$ is obtained by setting $n=0$)
\begin{equation}
\sqrt{  ( n + c)^2 + a^2}
   + \sqrt{ (n - c)^2 + a^2
   }=2\mathcal{J}+\frac{4n^2+k^2(1+6u+u^2)}{4\mathcal{J}}+
   O({1\ov \mathcal{J}^3})
\end{equation}

\subsection{Evaluation  of $E_1$ }

The final expression for $E_1$ in \rf{e1} is thus given by the sum
of \rf{spheref},\rf{adsf},\rf{sfq} and \rf{sf}.
 Let us split $E_1$ into the contribution of
 zero ($n=0$)   and non-zero ($n > 1 $)  modes
 \be
 E_1 = E_1^{(0)}  + \bar E_1  \ , \ \ \ \ \ \ \
 E_1^{(0)}=
 { 1 \ov 2\k} ( 4 \nu  + 2\k  +  \w_0  - 8 \td w_0 ) \ , \la{zero}
 \ee
 \bea \la{nonz}
 \bar E_1=  &&{ 1 \ov \k}\sum_{n=1}^\infty
 \bigg( 4 \sqrt{n^2 + \nu^2}+ 2 \sqrt{n^2 + \k^2}
 + \ha  \sum^{4}_{I=1} sign(C_{I,B}^{(n)}) \om_{I,n} \nn\\
 &&\ \ \  - \ 4 \big[ \sqrt{  ( n + c)^2 + a^2}
   + \sqrt{ (n - c)^2 + a^2 }\ \big] \bigg)\ ,  \eea

It is easy to check that the  sum over $n$ is convergent. Using
the large $n$ asymptotics of \be \sqrt{n^2+\nu^2}=
|n|+\frac{\nu^2}{2|n|}+\dots\ , \ \ \
\sqrt{n^2+\k^2}=|n|+\fr{\k^2}{2|n|}+\dots\ ,
 \label{adsflargen}
 \ee
and of the 4 solutions of \rf{bosch}
\be
 \w_{I=1,3; n} =
 |n|\bigg( 1 \pm  { 1 \ov|n|}   \sqrt{(1+\r_1^2)(\sqrt{\k^2 + k^2}- k )^2 -\r_1^2\k^2}
 \;+\frac{\kappa^2}{2n^2 }+\dots \bigg) \ ,
 \ee
 \be
 \w_{I=2,4; n } =
 - |n|\bigg(  1 \mp  { 1 \ov|n|} \sqrt{(1+\r_1^2)(\sqrt{\k^2 + k^2} + k )^2 -\r_1^2\k^2}
 \;  + \frac{\kappa^2}{2n^2 }+\dots  \bigg) \ ,
 \ee
we find
 \be\la{fi}
\bar E_1 = { 1 \ov  \k}\sum^\infty_{n=1}
\bigg(   \big[ 8n+\fr{2 (\k^2+\nu^2)}{n}+\dots \big]  -
\big[ 8 n + {4 a^2 \ov n} + \dots \big]\bigg) \ . \ee
We used here the signs of $C_{I,B}^{(n)}$ for the above
expressions of frequencies $\w_{I=1,2,3,4; n }$, which are
$sign(C_{1,B}^{(n)})=sign(C_{3,B}^{(n)})=+1$,
$sign(C_{2,B}^{(n)})=sign(C_{4,B}^{(n)})=-1$.\foot{Not that
exactly the same signs of $C_{I,B}^{(n)}$ we found before for the
frequencies at large $\mathcal{J}.$ But this is not necessarily
true for any $\mathcal{J}$ and $n.$} Thus the divergent terms in
the sum over $n$ indeed cancel between the bosonic and fermionic
contributions:
 according to \rf{cav}  $a^2 =\ha ( \k^2 + \nu^2)$.

 \bigskip

We  are interested in the  large $\J$ expansion  of
 $E_1 (\J,\S,k)$
at fixed  winding number $k$ (or $m$) and fixed $ \S/\J$.
Let us look first at the zero-mode part $E_1^{(0)}$ \rf{zero}
which is known
explicitly. Expanding it at large $\J$ we find
\be \la{hz}
E_1^{(0)} =  -    {   k^2 u(1+ u) \ov 2\J^2}    +  O( {1 \ov \J^4}) \ ,
\ \ \ \ \ \ \ \ \ \ \ \ \   u = \soj \ . \ee
Surprisingly, this already coincides
with the result of \ci{kz}  for the subleading $1/J$ correction
in the 1-loop anomalous dimension \rf{sym} of the corresponding SYM
operator.\footnote{Similar observation applies to the
$SU(2)$ case with $J_{1}=J_{2}$ considered
 in \cite{fpt}: there the coefficient of the $1/ \J^2$ term
  in the
  zero-mode part of the sum  for
  the 1-loop energy also has the same form
 as the 1-loop gauge theory expression found in  \cite{lz}.
}
In general, one expects that $\bE_1$ in \rf{nonz}
has the following expansion
\bea\la{ex}
 \bE_1
 =\fr{f_2\left(u,k\right)}{\mj^2}+\fr{f_4(u,k)}{\mj^4}+\dots\ .
 \eea
To find  $f_2$
 we are supposed to first do the sum in \rf{nonz}
and then expand the result at large $\J$.
However, given that computing the sum explicitly
  for any $\J$
does not seem possible, one
may attempt to  first expand the
frequencies at large $\J$ and then do the sum over $n$,
separately for each $1/\J^k $ coefficient.
This   may  not be consistent in general: the procedures of
 expanding  in $1/\J$ and  summing over $n$
may not commute.\foot{Indeed,   in taking
$\J$ large  in the frequencies one assumes that $n$ is fixed,
but $n$ can take  arbitrarily large values in the sum.
 The  consistency of this procedure
 depends on how rapidly
 the expansion and the
 sum  are  converging: it is likely  that the sums over $n$
 in the coefficients of higher $1/\J^k$  terms become divergent
 at large $n$, despite the fact that, as we have seen above,
 the sum in $\bE_1$ is finite at fixed $\J$.
 Then in general  one may not be able
to  trust the  coefficient of the $1/\J^2$
  term obtained by ``first expanding,  then doing the sum''
  procedure: one will need to resum the (divergent) expansion in
  $1/\J$.}

  Applying this   procedure of first expanding the frequencies at large $1/\J$
 one  finds
  the expected $ 1/ \J^2$ asymptotics, and moreover,
   that, remarkably,   the coefficient of the $1/\J^2$ term
    is given by a {\it convergent} sum over $n$.
  Explicitly, expanding the bosonic and fermionic terms in    \rf{nonz}
  at large $\J $  we get
  \bea
&&  \bE_1=  \sum_{n=1}^\infty
 \bigg(   8 +
 \frac{1}{2\mathcal{J}^2}\big[7n^2 +2k^2 (1+u)(1-4u)
+n\sqrt{n^2+4k^2u(1+u)}\big]+O({ 1 \ov \mathcal{J}^4}) \bigg) \nn \\
&&- \sum_{n=1}^\infty  \bigg(  8 +
 \frac{1}{\mathcal{J}^2}\big[4n^2 +k^2 (1+u)(1-3u) \big]
 +O({ 1 \ov \mathcal{J}^4}) \bigg)
 =
\frac{\td f_2(u,k) }{\J^2} + O({ 1 \ov \mathcal{J}^4})\ .\la{onz}
\eea
 Here $\td f_2$ is  given by the  convergent series
 \be \la{nai}
\td f_2(u,k)=-  \ha \sum_{n=1}^{\infty} \big[ n^2 + 2k^2 u (1 + u)
- n\sqrt{n^2+4k^2u(1+u)}\ \big]
\ ,\ee
 and tilde indicates that this is
  a  ``naive'' value of $f_2$.
Note that $\td f_2$  is a continuous negative function of
$x= 2k^2 u (1 + u)$ which vanishes at $x=0$.
Numerical evaluation of the sum  in
   (\ref{nai})   with $n\leq 10^5$  gives the following values
   (for $k=1$)
   \begin{center}
\begin{tabular}{cccccccccccccccccccc}
 \hline
  $u$   & \vline       & 0.5 &\vline
             & 0.6 &\vline & 0.7 &\vline & 0.8 &\vline
             & 0.9 &\vline &  1 &\vline & 1.1 &\vline & 1.2 \\
 \hline
  $\tilde{f}_2\left(u,1\right)$ & \vline & -0.56  &\vline  & -0.85 &\vline
  & -1.22 &\vline   & -1.68 &\vline
               &  -2.23 &\vline  &  -2.88 &\vline & -3.65 &\vline & -4.53 \\
 \hline
\end{tabular}
\end{center}

 The coefficients  of higher $O({ 1/ \mathcal{J}^{2k}})$
 terms in the ``naive'' expansion \rf{onz}
 are found to be divergent, so the $1/\J$  expansion needs to be
 resummed,  and the correct value of $f_2$ need not
a priori  coincide with $\td f_2$.
Surprisingly,  the results of the
  direct numerical evaluation of $f_2$
 discussed below turn out to be essentially   the same  as the
 above results for  $\td f_2$. That means
 that resummation of the higher $1/\J^4, ...$ terms in \rf{onz}
 does not change the coefficient of the $1/\J^2$ term:
the ``naive'' coefficient $\td f_2$ is actually the correct one.

To  compute the sum over $n$ in   \rf{nonz}
numerically
at fixed $k,u={\S\ov \J}$ and  large $\mathcal{J}$
we have  first  simplified the parameters in the frequencies
(but not expanding full $\om_{I,n}$ at fixed $n$)
in (\ref{spheref}),(\ref{adsf}),(\ref{bosch}) and (\ref{sf})
using that
$m=-k u, \ \n^2=
  \mathcal{J}^2-k^2 u^2$ and that
  $\k$ has expansion given in \rf{ku}.
We  then  estimated
the  coefficient $f_2$ in \rf{ex}
 for various values of $u$   (for simplicity we
  set $k=1$ and considered $u$ of order 1) by
computing   the sum  over $n$  with  the upper limit
  $N=5000$ and  varying  $\J$  in the
 interval $50\leq \mathcal{J}\leq 1000$.\foot{As
  was already mentioned, the solution is stable
  for large enough $\J$ (e.g., $\J > 50$)
   so the  frequencies here are real.}
 We  confirmed the form of the expansion  \rf{ex}
 with $f_2$ being {\it non-zero}.
  Its numerical values
 turned out to be the {\it same} as given in the table above for
 $\td f_2$. This demonstrates  that for the given
solution (and likely in some other similar cases where $\td f_2$ is given by a finite sum)
 the above
  procedure of first expanding  in large $\mathcal{J}$ and
 then doing  the sum is correct  at  order $1/\mathcal{J}^2$.\foot{It would
 be desirable  of course  to find an analytic proof of correctness
of this procedure.}

We conclude that the  expression for the 1-loop correction to the
string energy $E_{1}$, as obtained from
\rf{hz},\rf{ex},\rf{nai},  is given by
\begin{eqnarray}
E_{1}&=& -\frac{k^2 u(1+u)}{2\mathcal{J}^2} \nonumber \\
&-&\frac{1}{2\mathcal{J}^2} \sum_{n=1}^{\infty} \bigg[ n^2 + 2k^2 u
(1 + u) - n\sqrt{n^2+4k^2u(1+u)}\ \bigg]+O({1\ov \mathcal{J}^4}) \ .
\la{fina}
\end{eqnarray}

\section*{Acknowledgments }

We are grateful to  N. Beisert,  S. Frolov, V. Kazakov, M. Kruczenski,
 O. Lunin, A. Mikhailov, B. Stefanski, J. Russo  and K. Zarembo
for useful discussions and comments.
This  work  was supported  by the DOE
grant DE-FG02-91ER40690. The work of A.A.T. was also supported by
 the INTAS contract 03-51-6346
and RS Wolfson award.

\setcounter{footnote}{0}

\renewcommand{\theequation}{A.\arabic{equation}}
 \setcounter{equation}{0}
  \section*{Appendix A: Large $\J$ limit in  5-spin case
   }

Here we shall consider the expansion of the string action
in the case  when
the string may carry all
5 spins $(S_{1},S_{2},J_{1},J_{2},J_{3})$, thus
 generalizing  the discussion in \ci{st} and in sect. 2.
We shall still assume that
 isolating of the  ``fast'' variables is done in terms of
 the  3+3
 complex string coordinates in \rf{py},\rf{pya}, i.e.
  we shall define   $y$ and $\a$ as in \rf{uv}.
  This will not cover the cases of  more general
  string motions (like pulsations \ci{minna,mez})
    discussed in \ci{mik12,kt}.

 As follows from  \rf{ki},  the string Lagrangian takes the form
\begin{equation}
L=-\frac{1}{2}\sqrt{-g}g^{ab}\left(-D_{a}yD_{b}y+D_{a}\alpha
D_{b}\alpha+D_{a}V_{r}^{*}D_{b}V^{r}+D_{a}U_{i}^{*}D_{b}U^{i}\right)
\ . \label{nonconf1}
\end{equation}
As in the ($S,J)$-sector \rf{fixy},
we shall again fix the $U(1)$ symmetry
by choosing $y$ and $V_{r}$  so
that\  $y=t$ is the time coordinate
(that amounts to shifting angles in \rf{cartcoord1}
by $-t$).
 Two relevant combinations of charges are (cf. \rf{eee},\rf{charge})
\begin{equation}
\mathcal{E}-\mathcal{S}_{1}-\mathcal{S}_{2}=\int_{0}^{2\pi}
\frac{d\sigma}{2\pi}\ p_{y}\ , \quad \quad \mathcal{J}\equiv
\mathcal{J}_{1}+\mathcal{J}_{2}+\mathcal{J}_{3}=
\int_{0}^{2\pi}\frac{d\sigma}{2\pi}\ p_{\alpha}\ ,
\end{equation}
where
$
p_{\alpha}=-\sqrt{-g}g^{0a}D_{a}\alpha, \ \
p_{y}=p_{t}=-\sqrt{-g}g^{0a}D_{a} y\ .
$
Following \ci{kt},  we shall
introduce as in \rf{dual} the dual coordinate  $\tilde{\alpha}$
\begin{equation}
\sqrt{-g}g^{ab}D_{b}\alpha=-\epsilon^{ab}\partial_{b}\tilde{\alpha}\ ,
\label{dual1}
\end{equation}
so that after replacing $\a$ by $\td \a$
 and eliminating the 2d metric,
we get as in  \rf{hh} ($C_a= - i U^*_i \del_a U_i$, see  \rf{deriv})
\begin{equation}
L=\epsilon^{ab}C_{a}\partial_{b}\tilde{\alpha}-\sqrt{h} \ ,
\end{equation}
\begin{equation}
 h_{ab}=-D_{a}t
D_{b}t+\partial_{a}\tilde{\alpha}
\partial_{b} \tilde{\alpha} +D^*_{(a}V_{r}^{*}D_{b)}V^{r}
+D^*_{(a}U_{i}^{*}D_{b)}U^{i}\ .
\end{equation}
Choosing the same gauge as in \rf{gau}, i.e.
the static gauge in $t$ and $\td \a$,
\begin{equation}
t=\tau\ ,  \ \ \ \ \ \ \ \tilde{\alpha}=\mathcal{J}\sigma\ ,
\end{equation}
we find
$
L=C_{0}-\sqrt{h}$
with:
\begin{eqnarray}
h&=&|(-B_{1}^2+\mathcal{J}^2+|D_{1}V_{r}|^2+|D_{1}U_{i}|^2)
(|D_{0}V_{r}|^2+|D_{0}U_{i}|^2-(1+B_{0})^2)\nonumber\\
&-&\big[-B_{1}(1+B_{0})+\frac{1}{2}
(D^*_{0}V_{r}^{*}D_{1}V^{r}+D^*_{0}U_{i}^{*}D_{1}U^{i}+c.c.)\big]^2|
\end{eqnarray}
Expanding  at large $\J$ as in section 2 we get the action
\rf{ko} with \rf{LL1} replaced by
\begin{eqnarray}
\tilde{L}&=&-\mathcal{J}^2-i U^*_i \del_0 U_i
- i V^*_r \del_0 V^r
-\frac{1}{2}|D_{1}V_{r}|^2
-\frac{1}{2}|D_{1}U_{i}|^2\nonumber\\
&+&\frac{1}{8\mathcal{J}^2}[(|D_{1}V_{r}|^2+|D_{1}U_{i}|^2)^2
+4(|D_{0}V_{r}|^2+|D_{0}U_{i}|^2)-
4  i V^*_r \del_0 V^r    (|D_{1}V_{r}|^2+|D_{1}U_{i}|^2)\nonumber
\\
&+&4i V^*_r \del_1 V^r (D^*_{0}V_{r}^{*}D_{1}V^{r}
+D^*_{0}U_{i}^{*}D_{1}U^{i}+c.c.)] + O({ 1 \ov \mathcal{J}^4})\ .
\la{kp}
\end{eqnarray}
Finally, we can eliminate the time derivatives
from the ${ 1 / \mathcal{J}^2}$ term using leading-order equations
and then obtain the 2d energy.
Matching it with the charge  $E-S_{1}-S_{2}$
we obtain the space-time energy.

As in the ($S,J$)-sector,
 any possible winding of a closed string  along the two angles of
 $AdS_5$ is assumed to be
absorbed into  $V_{1,2}$ ($t$ must be  single-valued),  but in
contrast to the case of the  ($S,J$)-sector,
 now
it is natural to  absorb any possible windings in the three
 $S^{5}$ angles into $U_i$ (see also \ci{kt}).
  Then we get (cf. \rf{win})
$
\int_{0}^{2\pi} \frac{d\sigma}{2\pi} \partial_{1}\alpha=0
$, and
 using (\ref{dual1}) we obtain an  additional constraint
 on the solutions
\begin{equation}
\int_{0}^{2\pi} \frac{d\sigma}{2\pi}
C_{1}=-\mathcal{J}\int_{0}^{2\pi}
\frac{d\sigma}{2\pi}\frac{h_{01}}{\sqrt{h}}
\end{equation}
where
\begin{equation}
h_{01}=-B_{1}(1+B_{0})+\frac{1}{2}(D^*_{0}V_{r}^{*}D_{1}V^{r}
+D^*_{0}U_{i}^{*}D_{1}U^{i}+c.c.)
\end{equation}
Rescaling
 $ \tau \rightarrow \mathcal{J}^2\tau\ , \
\partial_{0}\rightarrow \frac{1}{\mathcal{J}^2}\partial_{0}$ and
expanding  in  large $\mathcal{J}$ this becomes
(cf.\rf{constraint})
\begin{eqnarray}
- \int_{0}^{2\pi}\frac{d\sigma}{2\pi} \ i  U^*_i \del_1 U_i
 &=&\int_{0}^{2\pi}  \frac{d\sigma}{2\pi}   \ i V^*_r \del_1 V^r
 - \frac{1}{2\mathcal{J}^2}\int_{0}^{2\pi}\frac{d\sigma}{2\pi}
\bigg[ i V^*_r \del_1 V^r  (|D_{1}V^{r}|^2+|D_{1}U_{i}|^2)
\nn \\
&+&(D^*_{0}V_{r}^{*}D_{1}V^{r}
+D^*_{0}U_{i}^{*}D_{1}U^{i}+c.c.)\bigg]+O({1 \ov \mathcal{J}^4})\ .
\end{eqnarray}
One  can again eliminate the time derivatives  here using
the equations of motion.

\renewcommand{\theequation}{B.\arabic{equation}}
 \setcounter{equation}{0}
  \section*{Appendix B:\ \  More on fermionic frequencies
 }

The fermionic operator  given in (\ref{dfo}) can be simplified
(as in \ci{ft3})  by  performing
appropriate rotations in the (36) and (06)-planes:
 \bea
 S_{36} S_{06}=e^{-\fr12 p\G_3\G_6}e^{-\fr12 q \G_0\G_6}\ , \ \ \ \ \
 \sin p=\fr{m}{\sqrt{m^2+k^2\r_1^2}}\ ,
 \ \ \ \ \ \sinh q  = -{w \ov k\r_1 } \ .  \label{spcp}
 \eea
Then  rotated  $D_F$ becomes
(we  rescale it  by an overall factor
$\sqrt{{\textstyle \ha}(\k^2-\n^2)}$)
\foot{
Some useful relations are
$ w\cos p-\varpi\r_1 \sin p=\fr{w}{k\r_1}\sqrt{m^2+k^2\r_1^2}$,
$m^2+k^2\r_1^2=\fr12 (\k^2-\n^2)$,

 $\k\r_0\sinh q+(w\cos p-\w\r_1 \sin p)\cosh q=0$. }
 \bea
&&D'_F=\G_0
 \bigg[\pa_\t  +  \fr{\r_0\r_1 k^2}
 {\sqrt{2(\k^2-\n^2)} }
 \G_{10}
  -\fr{k\varpi \r_0\r_1}
  {\sqrt{2(\k^2-\n^2)} }\G_{13}
  + \fr{\k
  m}{\varpi}\fr{\varpi^2-w^2}{\k^2-\n^2}\G_{16}
  \bigg]\nn\\
 && - \G_3\bigg[\pa_\s -
  \fr{\r_0}{\r_1}\fr{mw}{\sqrt{2(\k^2-\n^2)}}\G_{10}
  -\fr{k^2\r_0\r_1}{\sqrt{2(\k^2-\n^2)}}\G_{13}
  +  \fr{k\k m\r_0^2}{{\k^2-\n^2}}\G_{16}
  \bigg]
 \nn\\
&&+ { 2k\k \r_0\r_1 \ov \sqrt{ 2(\k^2-\n^2)}  }
\G_{124}\ .
 \eea
To find 16 characteristic frequencies we replace $\del_\tau$ by $i \om$ and
$\del_\s$ by $i n $ and solve  the equation
det$D'_F$=0 for $\om$.
To simplify the problem  may first restrict
to the subspaces
$\G_{24}\th=\pm i\th $ ($\G_{24}$ commutes
 with other matrices
in $D'_F$). Then
 \bea
 {\tilde D_F}
 =    i\w \G_0-i\,n\,\G_3 \pm  i a  \G_{1}
 +c \G_{016}  +   d \G_{136} \ ,
 \label{df}
 \eea
where
\bea
 a=\fr{\sqrt{2}k\k \r_0\r_1 }{\sqrt{\k^2-\n^2}} \; , \;\;\;\;\;
c=\fr{\k m}{\varpi}\fr{\varpi^2-w^2}{\k^2-\n^2}\;,\;\;\;\;\;\;\;
  \;\; d= \fr{k\k m\r_0^2}{{\k^2-\n^2}}\ .
 \eea
Multiplying \rf{df}  by $\G_0$  we can further restrict to
the subspaces
$
 \G_{0136}\th=\pm  i \th\
$ ($ \G_{0136}$ commutes with $\G_{24}$).
The sign degeneracy leads to simple degeneracy in
 the frequencies
and  we finish with the  following
characteristic equation
 \bea
 \w^2 \pm 2d\,\w - a^2 - c^2 + d^2 \pm  2 c\,n -n^2 =0\ .
 \eea
Its solution is
 \bea
\w =    \pm \sqrt{ (n \pm  c )^2  + a^2  } \ \pm d
 \ . \la{hi} \eea
There is also an  extra factor of 2 degeneracy
  making up the total  16 of fermionic  frequencies in \rf{e1}.
  One can show that $ a^2 + c^2 $ is always bigger than $d^2$
  and that $ a^2 > d^2 $  for $\J > k$
  (which is the large $\J$ region  we are interested in).

We observe that for any $n$ half of the fermionic frequencies
  are positive and half negative. According to \cite{blau}, we expect
 that  in this case
 $sign(C_{I,F}^{(n)})\omega_{I,n}^{F}=|\omega_{I,n}^{F}|.$
  As one can check,  this is indeed true here. Then taking into account
  that the square root term
  in the absolute
  value of $\omega$ in
   \rf{hi}  is bigger than $d$,\foot{Note that this is true for
   any $\J$ in the large  $n$ region  relevant
   for checking the finiteness of the sum in \rf{fi}
   which should hold for any value of $\J$.}
    we are led   to \rf{sf}.


\renewcommand{\theequation}{C.\arabic{equation}}
 \setcounter{equation}{0}
  \section*{Appendix C:  Landau-Lifshitz  frequencies
   }

The characteristic frequencies
obtained from the Landau-Lifshitz Lagrangian \rf{LL4} should
 form a subset of  all
bosonic characteristic frequencies expanded at large $\mathcal{J}$.
We can obtain the quadratic fluctuation
Lagrangian around the ($S,J)$ solution  discussed in section 3
  by introducing the fluctuations $
\eta\rightarrow \eta+\tilde{\eta},\ \rho\rightarrow
\rho_{0}+\tilde{\rho}$ and expanding
 to quadratic order in  $\tilde{\eta}, \tilde{\rho}
$. We then find  that the
characteristic frequencies are  given by
\begin{equation}
\omega_{\pm}=\frac{1}{2\mathcal{J}}\left[2nk(1+2\r_{1}^2)\pm
n\sqrt{n^2+4k^2\r_{1}^2(1+\r_{1}^2)}\right] \ , \label{freq}
\end{equation}
These frequencies should be seen from the Bethe ansatz as in
\cite{bmsz}.

Note, however, that the above frequencies differ from the
expansion of the full string frequencies in \rf{ol} by the term
$\frac{nk\r_{1}^2}{\mathcal{J}}. $\foot{The 1-loop energy 
computed  by summing 
these frequencies is still the same: the term proportional to 
$n$, i.e. $\frac{nk(1+\r_{1}^2)}{\mathcal{J}}$, does not
contribute when taking the sum over $n$ due to the weights
$C_{I}^{(n)}$ with which the frequencies appear in the sum
like  \rf{e}.} The
resolution to this puzzle is that one has to match carefully the
solution of the Landau-Lifshitz model (obtained from string theory
action using uniform gauge) and the string solution obtained in
the conformal gauge. Let $(\tau_{u},\sigma_{u})$ be the
world-sheet
 coordinates of the  uniform
gauge, and $(\tau_{c},\sigma_{c})$ the coordinates of the
conformal gauge. For the circular solution \rf{sol} in the
conformal gauge one has  $\eta=\phi_{1}-t=\rm w
\tau_{c}+k\sigma_{c}-\kappa \tau_{c}$, so that  expanding  at
large $\mathcal{J}$ we get
\begin{equation}
\eta=\frac{k^2}{2\mathcal{J}}\tau_{c}+k\sigma_{c} \ .   \label{solc}
\end{equation}
In the uniform gauge we have (here we have rescaled \rf{LL4} by
$\tau \rightarrow \mathcal{J}\tau$ giving
$t=\mathcal{J}\tau,\alpha=\mathcal{J}\tau+m\sigma$, as appropriate
for comparison to the circular solution in conformal gauge)
\begin{equation}
\eta=\frac{k^2(1+2\r_{1}^2)}{2\mathcal{J}}\tau_{u}+k\sigma_{u} \ .
\end{equation}
We see that the above expressions match when the two gauges are
related by:
\begin{equation}
\sigma_{u}=\sigma_{c}-\frac{k^2\r_{1}^2}{\mathcal{J}}\tau_{c}\ ,
\ \ \ \ \ 
\quad \tau_{u}=\tau_{c} \ . 
\end{equation}
Given that  the fluctuations are proportional to
$e^{i\omega\tau+in\sigma}$  this change  would 
produce the  following  frequencies in the conformal gauge
\begin{equation}
\omega_{\pm}=\frac{1}{2\mathcal{J}}\left[2nk(1+\r_{1}^2)\pm
n\sqrt{n^2+4k^2\r_{1}^2(1+\r_{1}^2)}\right] \ . \label{freq}
\end{equation}
These are indeed the same as the frequencies 
obtained directly  by expanding the 
string theory  frequencies in \rf{ol}.


\end{document}